\documentstyle[aps,eqsecnum,psfig]{revtex}

\begin{document}

\draft

\title{Parquet solution for a flat Fermi surface}

\author{Anatoley T.~Zheleznyak\cite{Zheleznyak} and Victor
M.~Yakovenko\cite{Yakovenko} }

\address{Department of Physics and Center for Superconductivity
Research, University of Maryland, College Park, MD 20742}

\author{Igor E.~Dzyaloshinskii}

\address{Department of Physics, University of California, Irvine, CA
92717}

\date{\bf cond-mat/9609118, September 12, 1996}

\maketitle

\begin{abstract}

   We study instabilities occurring in the electron system whose Fermi
surface has flat regions on its opposite sides.  Such a Fermi surface
resembles Fermi surfaces of some high-$T_c$ superconductors.  In the
framework of the parquet approximation, we classify possible
instabilities and derive renormalization-group equations that
determine the evolution of corresponding susceptibilities with
decreasing temperature.  Numerical solutions of the parquet equations
are found to be in qualitative agreement with a ladder approximation.
For the repulsive Hubbard interaction, the antiferromagnetic
(spin-density-wave) instability dominates, but when the Fermi surface
is not perfectly flat, the $d$-wave superconducting instability takes
over.

\end{abstract}


\section{Introduction}
\label{sec:intro}

   An interacting electron gas in one dimension has many unusual
properties, such as the spin-charge separation, the power law of
correlation functions, and the linear dependence of the electron
relaxation rate on temperature and frequency (see Ref.\
\cite{Firsov85} for a review).  These one-dimensional (1D) results are
well established, in many cases exactly, by applying a variety of
mathematical methods including the Bethe Ansatz, the bosonization, and
the parquet, or the renormalization group. To distinguish the exotic
behavior of the 1D electron gas from a conventional Fermi-liquid
behavior, Haldane introduced a concept of the so-called Luttinger
liquid \cite{Haldane81}.

   The discovery of high-$T_c$ superconductivity renewed interest in
the Luttinger-liquid concept.  Anderson suggested that a
two-dimensional (2D) electron gas behaves like the 1D Luttinger
liquid, rather than a conventional Fermi liquid \cite{Anderson92}.  It
is difficult to verify this claim rigorously, because the methods that
prove the existence of the Luttinger liquid in 1D cannot be applied
directly to higher dimensions.  The Bethe Ansatz construction does not
work in higher dimensions.  The bosonization in higher dimensions
\cite{Haldane92,Khveshchenko93a,Khveshchenko94b,Marston93,Marston,Fradkin,LiYM95,Kopietz95}
converts a system of interacting electrons into a set of harmonic
oscillators representing the electron density modes. This procedure
replaces the exact $W_\infty$ commutation relations
\cite{Khveshchenko94b} with approximate boson commutators, which is a
questionable, uncontrolled approximation. On the other hand, the
parquet method, although not being as exact as the two other methods,
has the advantage of being formulated as a certain selection rule
within a standard many-body diagram technique; thus, it can be applied
to higher dimensions rather straightforwardly. The parquet method has
much in common with the renormalization-group treatment of Fermi
liquids \cite{Shankar94}.

   The 1D electron gas has two types of potential instabilities: the
superconducting and the density-wave, which manifest themselves
through logarithmic divergences of the corresponding one-loop
susceptibilities with decreasing temperature.  Within the parquet
approach, a sum of an infinite series of diagrams, obtained by adding
and inserting the two basic one-loop diagrams into each other, is
calculated by solving a system of nonlinear differential equations,
which are nothing but the renormalization-group equations
\cite{Solyom79}.  This procedure was developed for the first time for
meson scattering \cite{Diatlov57} and later was successfully applied
to the 1D electron gas \cite{Bychkov66,Dzyaloshinskii72a}, as well as
to the Kondo problem \cite{Abrikosov} and the X-ray absorption edge
problem \cite{Nozieres69a}.  By considering both the superconducting
and the density-wave instabilities on equal footing and adequately
treating their competition, the parquet approximation differs from a
conventional ladder (or mean-field) approximation, commonly applied in
higher dimensions, where only one instability is taken into account.
Under certain conditions in the 1D case, the superconducting and
density-wave instabilities may cancel each other, giving rise to a
non-trivial metallic ground state at zero temperature, namely the
Luttinger liquid.  In this case, the parquet derivation shows that the
electron correlation functions have a power-law structure, which is
one of the characteristic properties of the Luttinger liquid
\cite{Dzyaloshinskii72a,Larkin73}.  One may conclude that the
competition between the superconducting and density-wave instabilities
is an important ingredient of the Luttinger liquid theory.

   In a generic higher-dimensional case, where density-wave
instability does not exist or does not couple to superconducting
instability because of corrugation of the Fermi surface, the parquet
approach is not relevant.  Nevertheless, there are a number of
higher-dimensional models where the parquet is applicable and produces
nontrivial results.  These include the models of multiple chains
without single-electron hopping \cite{Gorkov74} and with
single-electron hopping but in a magnetic field \cite{Yakovenko87}, as
well as the model of an isotropic electron gas in a strong magnetic
field \cite{Brazovskii71,Yakovenko93a}.  In all of these models, the
electron dispersion law is 1D, which permits to apply the parquet
method; at the same time, the interaction between electrons is
higher-dimensional, which makes a nontrivial difference from the
purely 1D case.  The particular version of the parquet method used in
these cases is sometimes called the ``fast'' parquet, because, in
addition to a ``slow'', renormalization-group variable, the parquet
equations acquire supplementary, ``fast'' variables, which label
multiple electron states of the same energy.

   Taking into account these considerations, it seems natural to start
exploring a possibility of the Luttinger liquid behavior in higher
dimensions by considering a model that combines 1D and
higher-dimensional features.  This is the model of an electron gas
whose Fermi surface has flat regions on its opposite sides.  The
flatness means that within these regions the electron dispersion law
is 1D: The electron energy depends only on the one component of
momentum that is normal to the flat section.  On the other hand, the
size of the flat regions is finite, and that property differentiates
the model from a purely 1D model, where the size is infinite, since
nothing depends on the momenta perpendicular to the direction of a 1D
chain.  A particular case of the considered model is one where the 2D
Fermi surface has a square shape.  This model describes 2D electrons
on a square lattice with the nearest-neighbor hopping at the half
filling.  It is a simplest model of the high-$T_c$ superconductors.

   The model has already attracted the attention of
theorists. Virosztek and Ruvalds studied the ``nested Fermi liquid''
problem within a ladder or mean-field approximation
\cite{Ruvalds90,Ruvalds95}.  Taking into account the 1D experience,
this approach may be considered questionable, because it does not
treat properly the competition between the superconducting and the
density-wave channels.  Houghton and Marston \cite{Marston93} mapped
the flat parts of the Fermi surface onto discrete points. Such an
oversimplification makes all scattering processes within the flat
portion equivalent and artificially enhances the electron interaction.
Mattis \cite{Mattis87} and Hlubina \cite{Hlubina94} used the
bosonization to treat the interaction between the electron density
modes and claimed to solve the model exactly. However, mapping of the
flat Fermi surface onto quantum chains and subsequent bosonization by
Luther \cite{Luther94} indicated that the treatment of Mattis and
Hlubina is insufficient, because the operators of backward and umklapp
scattering on different quantum chains require a consistent
renormalization-group treatment. Luther did not give solution to this
problems, as well as he missed the interaction between the electrons
located on four different quantum chains.

   In the present paper, we solve the model consistently, using the
fast parquet approach, where all possible instabilities occurring in
the electron system with the flat regions on the Fermi surface are
treated simultaneously.  This approach was applied to the problem
earlier \cite{Dzyaloshinskii72b} in order to explain the
antiferromagnetism of chromium.  In the present paper, we advance the
study further by including the order parameters of the odd symmetry,
missed in \cite{Dzyaloshinskii72b}, performing detailed numerical
calculations, and investigating the effect of a curvature of the Fermi
surface.  To simplify numerical calculations and to relate to the
high-$T_c$ superconductors, we consider the 2D case, although the
method can be straightforwardly generalized to higher dimensions as
well.

   We find that the presence of the boundaries of the flat portions of
the Fermi surface has a dramatic effect on the solutions of the
parquet equations.  Even if the initial vertex of interaction between
electrons does not depend on the momenta along the Fermi surface
(which are the ``fast'' variables), the vertex acquires a strong
dependence on these variables upon renormalization, which greatly
reduces the feedback coupling between the superconducting and
density-wave channels relative to the 1D case.  Instead of the two
channels canceling each other, the leading channel, which is the
spin-density-wave (SDW) in the case of the repulsive Hubbard
interaction, develops its own phase transition, inducing on the way a
considerable growth of the superconducting $d$-wave susceptibility.
At the same time, the feedback from the superconducting to the SDW
channel, very essential in the 1D case, is found negligible in the 2D
case.  These results are in qualitative agreement with the picture of
the antiferromagnetically-induced $d$-wave superconductivity, which
was developed within a ladder approximation for the flat Fermi surface
in \cite{Ruvalds95} and for a generic nested Hubbard model in
\cite{Scalapino}. Recent experiments strongly suggest that the
high-$T_c$ superconductivity is indeed of the $d$-wave type
\cite{d-wave}.  On the other hand, our results disagree with Refs.\
\cite{Mattis87,Hlubina94}. The origin of the discrepancy is that the
bosonization arbitrarily replaces the exact $W_\infty$ commutation
relations \cite{Khveshchenko94b} by approximate boson commutators;
thus, the renormalization of the electron-electron interaction, which
is an important part of the problem, becomes neglected.

   In addition to having the flat sides, the square Fermi surface also
has sharp corners, where the saddle points of the electron dispersion
law, which produce the van Hove singularity in the density of states,
are located.  The presence of the van Hove singularity at the Fermi
level enhances the divergence of the superconducting and density-wave
loops to the square of the temperature logarithm. The fast parquet
problem was formulated in this case in Ref.\ \cite{Dzyaloshinskii87a},
where the contribution from the flat sides, being less divergent than
the contribution from the saddle points, was neglected.  The present
paper completes the study by considering a Fermi surface with the flat
sides and rounded corners, that is, without saddle points at the Fermi
level.  Our physical conclusions for both models are in qualitative
agreement.

   As photoemission experiments \cite{ZXShen93} demonstrate (see also
\cite{Ruvalds95}), many of the high-$T_c$ superconductors indeed have
flat regions on their Fermi surfaces.  Hence, some of the results of
this paper may be applicable to these materials.  However, the primary
goal of our study is to elucidate general theoretical concepts rather
than to achieve detailed description of real materials.

   In order to distinguish the new features brought into the problems
by introducing higher dimensions, we present material in an inductive
manner.  In Sec.\ \ref{sec:spinless}, we recall the derivation of the
parquet equations in the simplest case of 1D spinless electrons.  In
Sec.\ \ref{sec:spin1D}, we generalize the procedure to the case of 1D
electrons with spin \cite{Bychkov66,Dzyaloshinskii72a}. Then, we
derive the parquet equations in the 2D case in Sec.\ \ref{sec:2D} and
solve them numerically in Sec.\ \ref{sec:numerical}.  The paper ends
with conclusions in Sec.\ \ref{sec:conclusion}.

\section{Parquet Equations for One-Dimensional Spinless Fermions}
\label{sec:spinless}

\begin{figure}
\centerline{\psfig{file=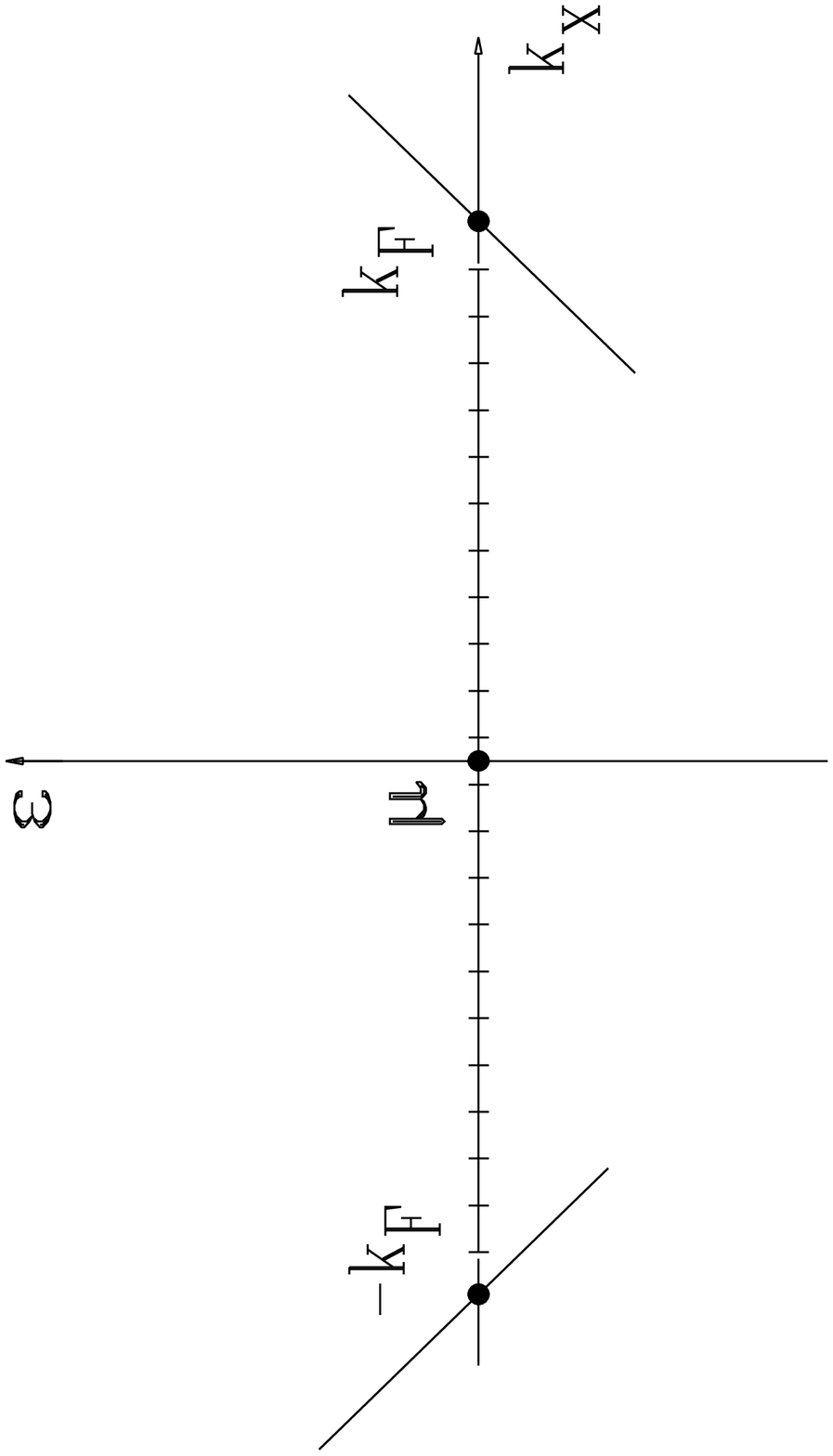,height=0.9\textheight}}
\caption{Dispersion law of 1D electrons.  The states in the shaded
range of the momentum $k_x$ are occupied by electrons.}
\label{fig:1D}
\end{figure}

   Let us consider a 1D electron gas with a Fermi energy $\mu$ and a
generic dispersion law $\varepsilon(k_x)$, where $\varepsilon$ is the
energy and $k_x$ is the momentum of the electrons.  As shown in Fig.\
\ref{fig:1D}, the Fermi surface of this system consists of two points
located at $k_x=\pm k_F$, where $k_F$ is the Fermi momentum.  Assuming
that the two points are well separated, let us treat the electrons
whose momenta are close to $\pm k_F$ as two independent species and
label them with the index $\pm$.  In the vicinity of the Fermi energy,
the dispersion laws of these electrons can be linearized:
\begin{equation}
   \varepsilon_{\pm}(k_x) = \pm v_F k_x ,
\label{eps}
\end{equation}
where the momenta $k_x$ are counted from the respective Fermi points
$\pm k_F$ for the two species of the electrons, $\pm v_F$ are the
corresponding Fermi velocities, and the energy $\varepsilon$ is
counted from the chemical potential $\mu$. 

   First, let us consider the simplest case of electrons without spin.
The bare Hamiltonian of the interaction between the $\pm$ electrons,
$\hat{H}_{\rm int}$, can be written as
\begin{equation}
   \hat{H}_{\rm int}= g \int\frac{dk_x^{(1)}}{2\pi}
   \frac{dk_x^{(2)}}{2\pi}\frac{dk_x^{(3)}}{2\pi}
   \hat{\psi}^+_+(k_x^{(1)}+k_x^{(2)}-k_x^{(3)})
   \hat{\psi}^+_-(k_x^{(3)}) \hat{\psi}_-(k_x^{(2)})
   \hat{\psi}_+(k_x^{(1)}),
\label{Interaction:spinless}
\end{equation}
where $g$ is the bare vertex of interaction, and the operators
$\hat{\psi}^+_\pm$ and $\hat{\psi}_\pm$ create and destroy the $\pm$
electrons.

\begin{figure}
\centerline{\psfig{file=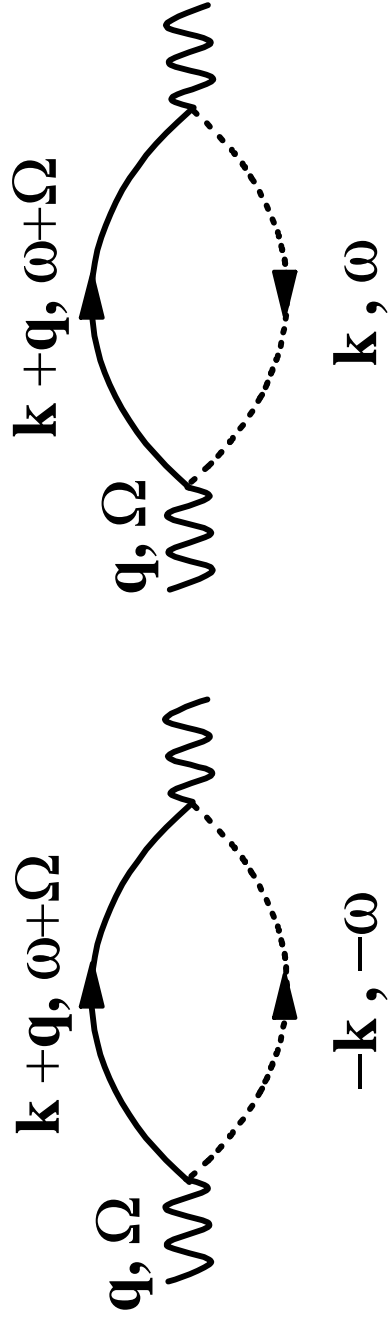,height=0.9\textheight}}
\caption{Bare superconducting and density-wave susceptibilities.  The
solid and dashed lines represent the Green functions of the $+$ and
$-$ electrons. The wavy lines represent incoming momentum and energy.}
\label{fig:loops}
\end{figure}

   The tendencies toward the superconducting or density-wave ($2k_F$)
instabilities in the system are reflected by the logarithmic
divergences of the two one-loop diagrams shown in Fig.\
\ref{fig:loops}, where the solid and dashed lines represent the Green
functions $G_+$ and $G_-$ of the $+$ and $-$ electrons, respectively.
The two diagrams in Fig.\ \ref{fig:loops} differ in the mutual
orientation of the arrows in the loops.  In the Matsubara technique,
the integration of the Green functions over the internal momentum
$k_x$ and energy $\omega_n$ produces the following expressions for the
two diagrams:
\begin{eqnarray}
   &&\pm T\sum_n\int\frac{dk_x}{2\pi}
   G_-(\mp\omega_n,\mp k_x)G_+(\omega_n+\Omega_n,k_x+q_x)
\nonumber \\
   &&=-T\sum_n\int\frac{dk_x}{2\pi}\frac{1}
   {(i\omega_n+v_Fk_x)(i\omega_n+i\Omega_m -v_F(k_x+q_x))}
\nonumber \\
   &&\approx\frac{1}{2\pi v_F}\ln\left(\frac{\mu}
   { \max\{T,|v_Fq_x|,|\Omega_m|\} } \right) \equiv \xi,
\label{1loop}
\end{eqnarray}
where the upper sign corresponds to the superconducting and the lower
to the density-wave susceptibility.  In Eq.\ (\ref{1loop}), $T$ is the
temperature, $\Omega_m$ is the external energy passing through the
loop, and $q_x$ is the external momentum for the superconducting loop
and the deviation from $2k_F$ for the density-wave loop.  With
logarithmic accuracy, the value of the integral (\ref{1loop}) is
determined by the upper and lower cutoffs of the logarithmic
divergence.  In Eq.\ (\ref{1loop}), the upper and lower cutoffs are
written approximately, up to numerical coefficients of the order of
unity, whose logarithms are small compared to $\xi\gg1$.  The variable
$\xi$, introduced by Eq.\ (\ref{1loop}), plays a very important role
in the paper.  Since $\xi$ is the logarithm of the infrared cutoff,
the increase of $\xi$ represents renormalization toward low
temperature and energy.

\begin{figure}
\centerline{\psfig{file=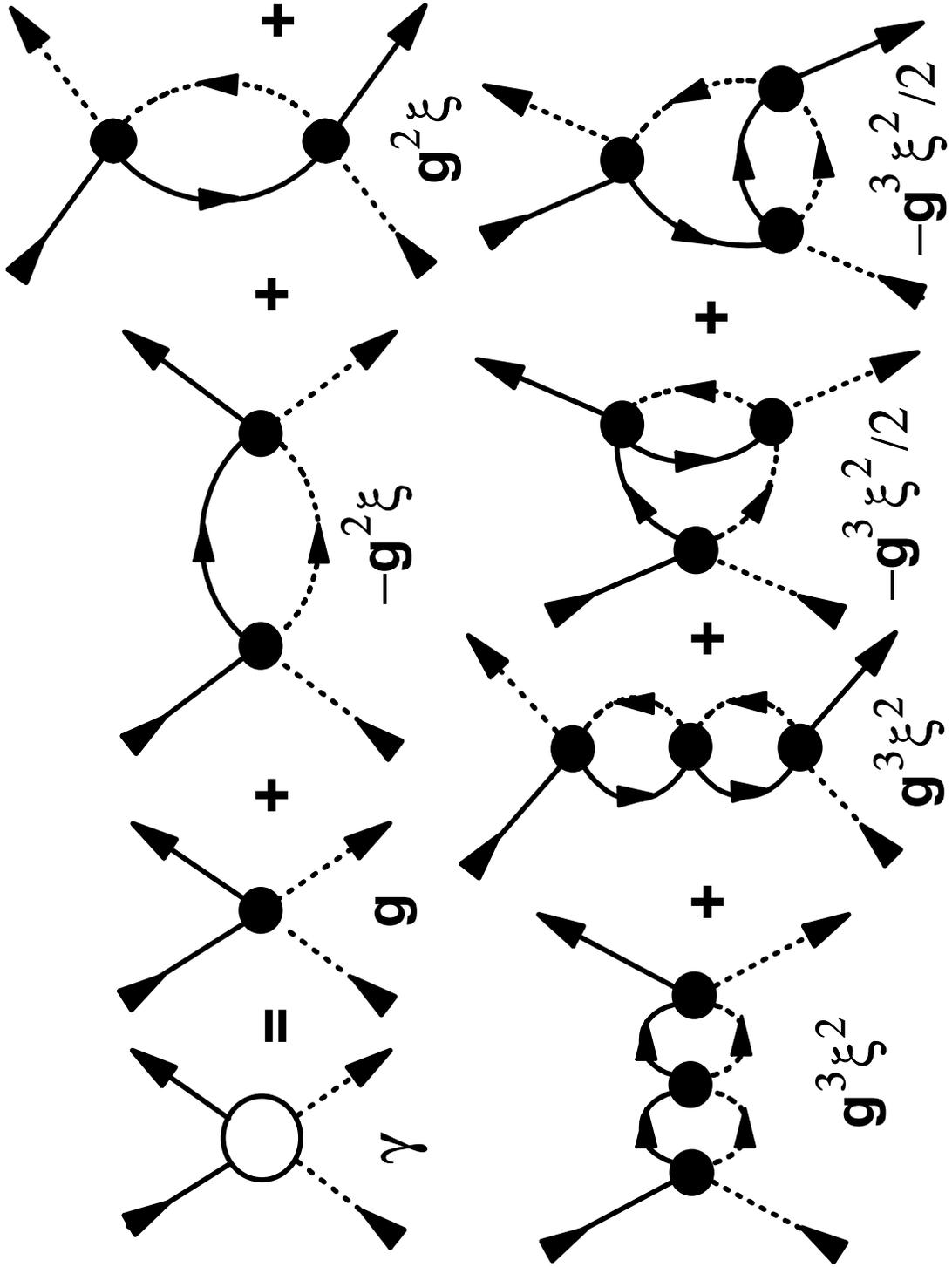,height=0.9\textheight}}
\caption{Some parquet corrections to the vertex of interaction between
electrons, $\gamma$, which is shown as a circle.  The dots represent the
bare interaction vertex $g$.  The expressions beneath the diagrams
represent the values of the corresponding diagrams.}
\label{fig:sample}
\end{figure}

   The two primary diagrams of Fig.\ \ref{fig:loops} generate
higher-order corrections to the vertex of interaction between
electrons, $\gamma$, as illustrated in Fig.\ \ref{fig:sample}.  In
this Figure, the dots represent the bare interaction vertex $g$,
whereas the renormalized vertex $\gamma$ is shown as a circle.  The
one-loop diagrams in Fig.\ \ref{fig:sample} are the same as in Fig.\
\ref{fig:loops}.  The first two two-loop diagrams in Fig.\
\ref{fig:sample} are obtained by repeating the same loop twice in a
ladder manner.  The last two diagrams are obtained by inserting one
loop into the other and represent coupling between the two channels.
The diagrams obtained by repeatedly adding and inserting the two basic
diagrams of Fig.\ \ref{fig:loops} in all possible ways are called the
parquet diagrams.  The ladder diagrams, where only the addition, but
not the insertion of the loops is allowed, represent a subset of the
more general set of the parquet diagrams.  Selection of the parquet
diagrams is justified, because, as one can check calculating the
diagrams in Fig.\ \ref{fig:sample}, they form a series with the
expansion parameter $g\xi$: $\gamma=g\sum_{n=0}^\infty a_n(g\xi)^n$.
If the bare interaction vertex $g$ is small and the temperature is
sufficiently low, so that $\xi(T)$ is big, one can argue
\cite{Diatlov57,Bychkov66,Dzyaloshinskii72a} that nonparquet diagrams
may be neglected, because their expansion parameter $g$ is small
compared to the parquet expansion parameter $g\xi$.

\begin{figure}
\centerline{\psfig{file=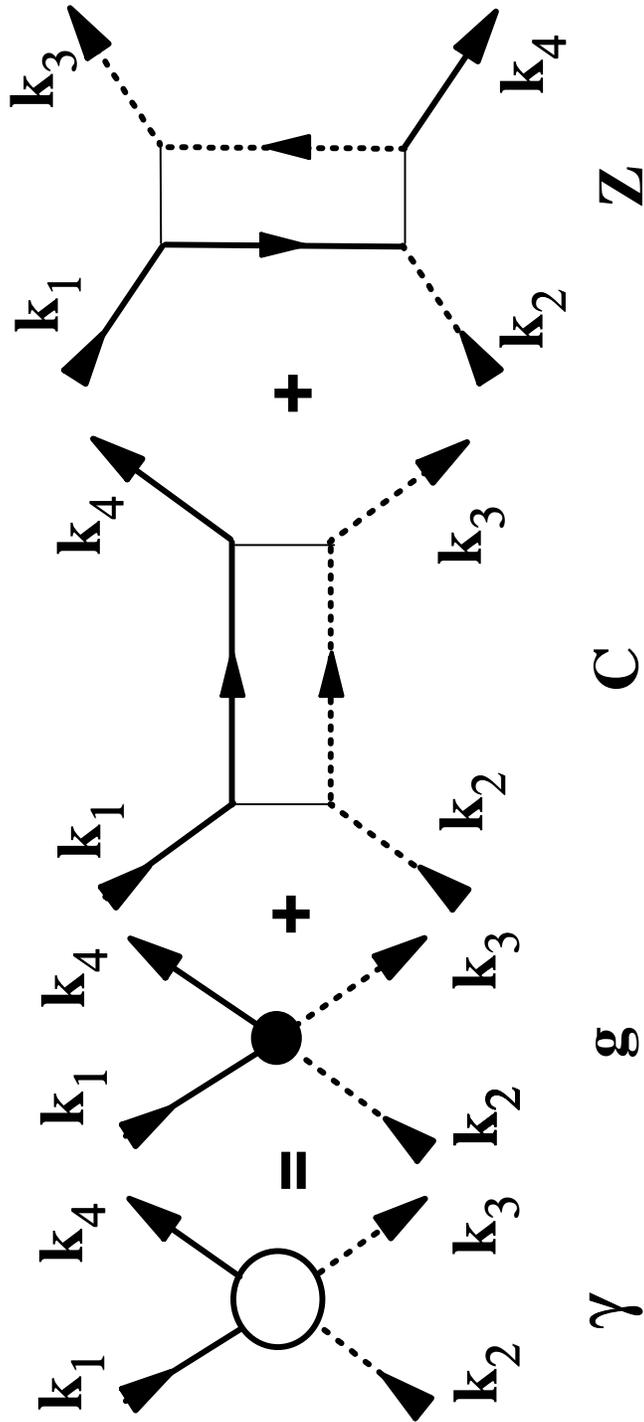,height=0.9\textheight}}
\caption{Decomposition of the interaction vertex $\gamma$, shown as a
circle, into superconducting and density-wave bricks, shown as
rectangles, in the spinless case.}
\label{fig:SpinlessVertex}
\end{figure}

   Every diagram in Fig.\ \ref{fig:sample}, except the bare vertex $g$,
can be divided into two disconnected pieces by cutting one solid and one
dashed line, the arrows of the cut lines being either parallel or
antiparallel.  The sum of those diagrams where the arrows of the cut
lines are parallel (antiparallel) is called the superconducting
(density-wave) ``brick''.  Thus, the vertex $\gamma$ can be decomposed
into the bare vertex $g$, the superconducting brick $C$, and the
density-wave brick $Z$:
\begin{equation}
   \gamma=g+C+Z.
\label{vertex:spinless}
\end{equation}
Eq.\ (\ref{vertex:spinless}) is illustrated in Fig.\
\ref{fig:SpinlessVertex}, where the bricks are represented as
rectangles whose long sides,  one being a solid and another a dashed
line,  represent the lines to be cut.

\begin{figure}
\centerline{\psfig{file=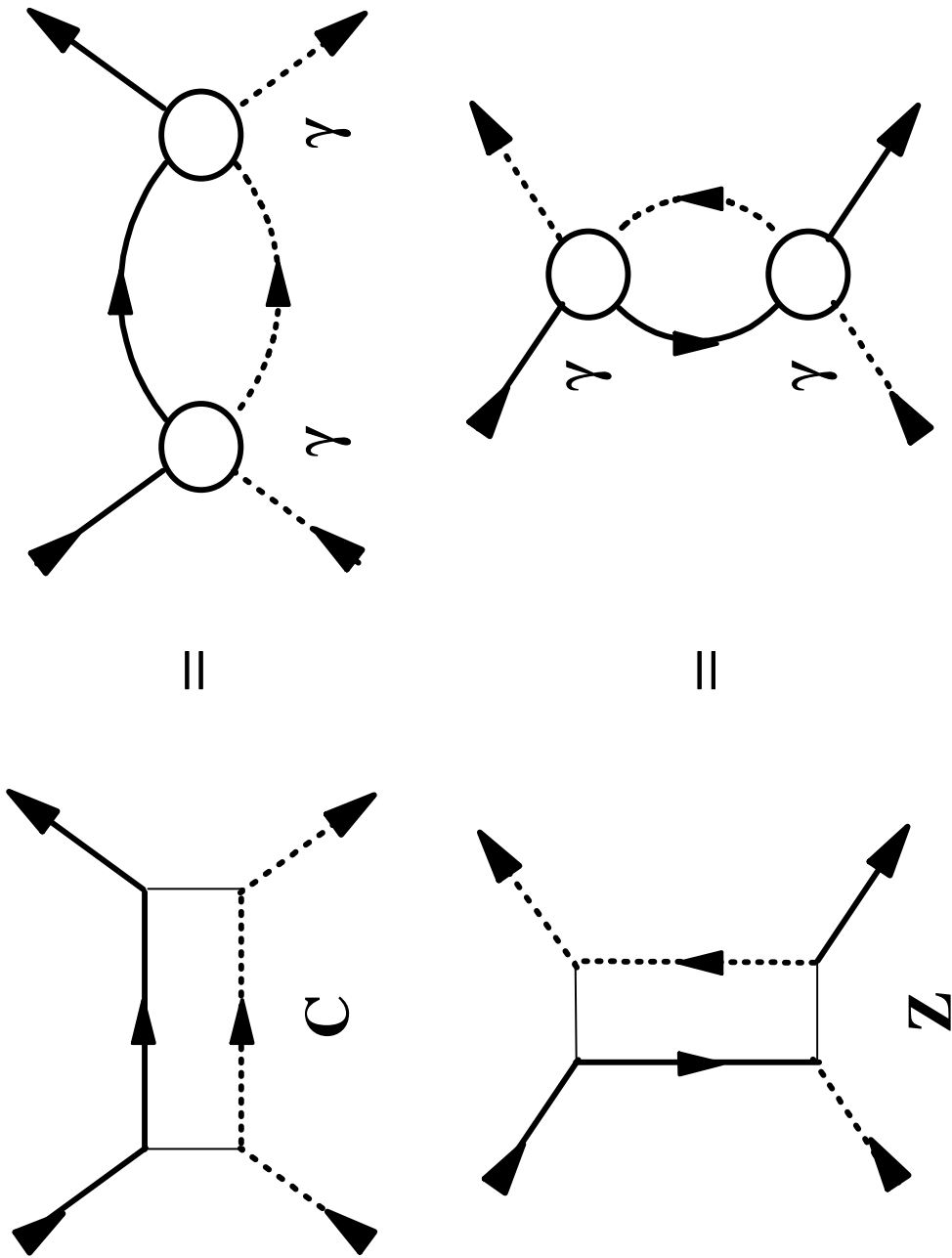,height=0.9\textheight}}
\caption{Parquet equations for the bricks in the spinless case.}
\label{fig:SpinlessBricks}
\end{figure}

   In a general case, the vertices and the bricks depend on the
energies and momenta
($\omega_1,\:\omega_2,\:\omega_3,\:v_Fk_x^{(1)},\:v_Fk_x^{(2)}$, and
$v_Fk_x^{(3)}$) of all incoming and outgoing electrons.  Equations for
the bricks can be found in closed form in the case where all their
arguments are approximately equal within the logarithmic accuracy,
that is, the ratios of the arguments and of their linear combinations
are of the order of unity
\cite{Diatlov57,Bychkov66,Dzyaloshinskii72a}.  Practically, this means
that all vertices and bricks are considered to be functions of the
single renormalization-group variable $\xi$, defined in Eq.\
(\ref{1loop}).  It was proved in \cite{Diatlov57} that the two pieces
obtained by cutting a brick are the full vertices of interaction, as
illustrated graphically in Fig.\ \ref{fig:SpinlessBricks}.
Analytically, the equations for the bricks are
\begin{mathletters}%
\label{integral}
\begin{eqnarray}
   C(\xi)&=&-\int_0^\xi d\zeta\,\gamma(\zeta)\gamma(\zeta),
\label{C}\\
   Z(\xi)&=&\int_0^\xi d\zeta\,\gamma(\zeta)\gamma(\zeta).
\label{Z}
\end{eqnarray}
\end{mathletters}%
The two vertices $\gamma$ in the r.h.s.\ of Eqs.\ (\ref{integral})
represent the two pieces obtained from a brick by cutting, whereas the
integrals over $\zeta$ represent the two connecting Green functions
being integrated over the internal momentum and energy of the loop.
The value of the renormalized vertex $\gamma(\zeta)$ changes as the
integration over $\zeta$ progresses in Eqs.\ (\ref{integral}).  In
agreement with the standard rules of the diagram technique \cite{AGD},
a pair of the parallel (antiparallel) lines in Fig.\
\ref{fig:SpinlessBricks} produces a negative (positive) sign in the
r.h.s.\ of Eq.\ (\ref{C}) [(\ref{Z})].

   Eqs.\ (\ref{integral}) can be rewritten in differential,
renormalization-group form:
\begin{mathletters}%
\label{differential}
\begin{eqnarray}
   && \frac{dC(\xi)}{d\xi}=-\gamma(\xi)\gamma(\xi),
   \quad\quad C(\xi\!\!=\!\!0)=0; \\
   && \frac{dZ(\xi)}{d\xi}=\gamma(\xi)\gamma(\xi),
   \quad\quad Z(\xi\!\!=\!\!0)=0.
\end{eqnarray}
\end{mathletters}%
Combining Eqs.\ (\ref{differential}) with Eq.\
(\ref{vertex:spinless}),  we find the renormalization equation for the
full vertex $\gamma$:
\begin{mathletters}%
\label{RG:spinless}
\begin{eqnarray}
   && \frac{d\gamma(\xi)}{d\xi}=\gamma(\xi)\gamma(\xi)
   -\gamma(\xi)\gamma(\xi)=0,
\label{cancellation} \\
   && \gamma(\xi\!\!=\!\!0)=g.
\end{eqnarray}
\end{mathletters}%
We see that the two terms in the r.h.s.\ of Eq.\ (\ref{cancellation}),
representing the tendencies toward density-wave and superconducting
instabilities, exactly cancel each other.  In a ladder approximation,
where only one term is kept in the r.h.s., the result would be quite
different, because $\gamma(\xi)$ would diverge at a finite $\xi$
indicating an instability or generation of a pseudogap in the system.

   In order to study possible instabilities in the system, we need to
calculate corresponding generalized susceptibilities.  For that
purpose, let us add to the Hamiltonian of the system two fictitious
infinitesimal external fields $h_{\rm SC}$ and $h_{\rm DW}$ that
create the electron-electron and electron-hole pairs:
\begin{eqnarray}
   \hat{H}_{\rm ext}=\int\frac{dq_x}{2\pi}\frac{dk_x}{2\pi}&&\left[
   h_{\rm SC}(q_x)\,\hat{\psi}^+_-\left(\frac{q_x}{2}-k_x\right)
   \hat{\psi}^+_+\left(\frac{q_x}{2}+k_x\right) \right.
\nonumber \\
   &&\left.{}+h_{\rm DW}(q_x)\,\hat{\psi}^+_-\left(k_x+\frac{q_x}{2}\right)
   \hat{\psi}_+\left(k_x-\frac{q_x}{2}\right) + {\rm H.c.} \right].
\label{Hext}
\end{eqnarray}

   Now we need to introduce triangular vertices ${\cal T}_{\rm SC}$
and ${\cal T}_{\rm DW}$ that represent the response of the system to
the fields $h_{\rm SC}$ and $h_{\rm DW}$.  Following the same
procedure as in the derivation of the parquet equations for the bricks
\cite{Bychkov66,Dzyaloshinskii72a,Brazovskii71,Dzyaloshinskii72b}, we
find the parquet equations for the triangular vertices in graphic
form, as shown in Fig.\ \ref{fig:SpinlessTriangle}.  In that Figure,
the filled triangles represent the vertices ${\cal T}_{\rm SC}$ and
${\cal T}_{\rm DW}$, whereas the dots represent the fields $h_{\rm
SC}$ and $h_{\rm DW}$.  The circles, as in the other Figures,
represent the interaction vertex $\gamma$.  Analytically, these
equations can be written as differential equations with given initial
conditions:
\begin{mathletters}%
\label{triangular}
\begin{eqnarray}
   \frac{d{\cal T}_{\rm SC}(\xi)}{d\xi}=-\gamma(\xi)
      {\cal T}_{\rm SC}(\xi),
      &\quad\quad\quad& {\cal T}_{\rm SC}(0)=h_{\rm SC};  \\
   \frac{d{\cal T}_{\rm DW}(\xi)}{d\xi}=\gamma(\xi)
      {\cal T}_{\rm DW}(\xi),
      &\quad\quad\quad& {\cal T}_{\rm DW}(0)=h_{\rm DW}.
\end{eqnarray}
\end{mathletters}%
We will often refer to the triangular vertices ${\cal T}$ as the
``order parameters''.  Indeed, they are the superconducting and
density-wave order parameters induced in the system by the external
fields $h_{\rm SC}$ and $h_{\rm DW}$.  If, for a finite $h_i$ ($i$=SC,
DW), a vertex ${\cal T}_i(\xi)$, which is proportional to $h_i$,
diverges when $\xi\rightarrow\xi_c$, this indicates that a {\em
spontaneous} order parameter appears in the system, that is, the order
parameter may have a finite value even when the external field $h_i$
is zero.  The external fields are introduced here only as auxiliary
tools and are equal to zero in real systems.  We also note that the
two terms in the r.h.s.\ of Eq.\ (\ref{Hext}) are not Hermitially
self-conjugate; thus, the fields $h_i$ are the complex fields.
Consequently, the order parameters ${\cal T}_i(\xi)$ are also complex,
so, generally speaking, ${\cal T}$ and ${\cal T}^*$ do not coincide.
According to Eqs.\ (\ref{RG:spinless}), $\gamma(\xi)=g$, so Eqs.\
(\ref{triangular}) have the following solution:
\begin{mathletters}%
\label{triangular:solutions}
\begin{eqnarray}
   {\cal T}_{\rm SC}(\xi)&=&h_{\rm SC}\exp(-g\xi), \\
   {\cal T}_{\rm DW}(\xi)&=&h_{\rm DW}\exp(g\xi).
\end{eqnarray}
\end{mathletters}%

\begin{figure}
\centerline{\psfig{file=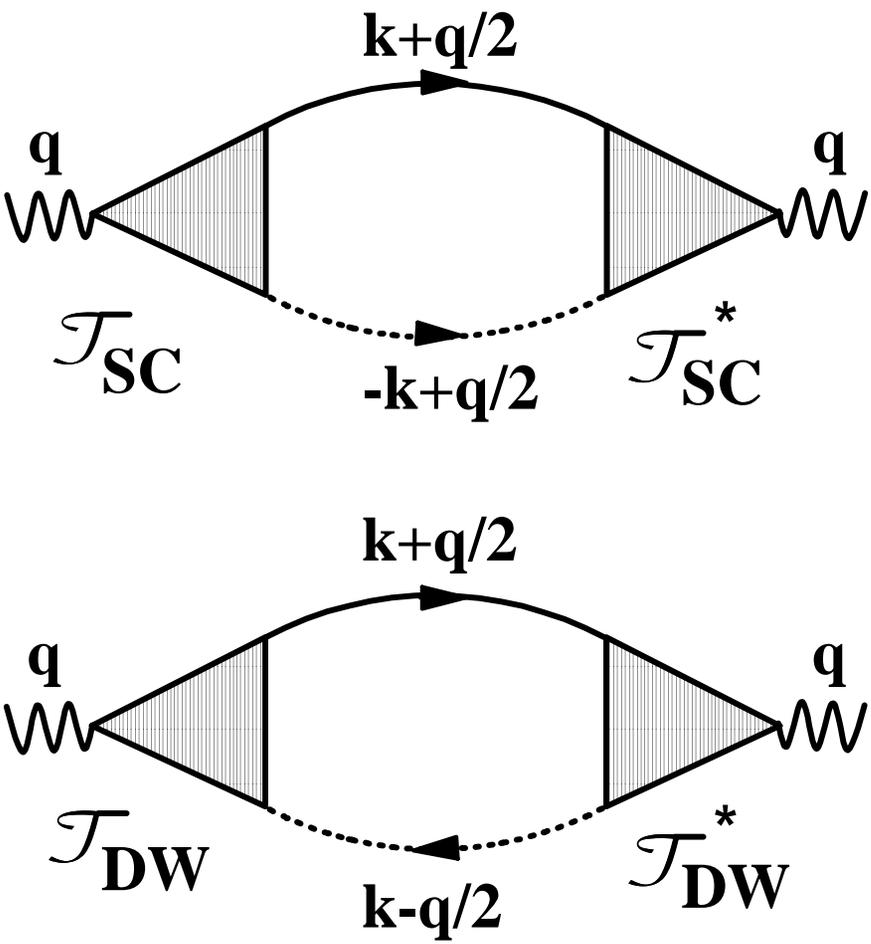,height=0.9\textheight}}
\caption{Parquet equations for the free energy corrections $F_{\rm
SC}$ and $F_{\rm DW}$ in the spinless case.}
\label{fig:SpinlessSusceptibility}
\end{figure}

   Now we can calculate the susceptibilities.  The lowest order
corrections to the free energy of the system due to the introduction
of the fields $h_{\rm SC}$ and $h_{\rm DW}$, $F_{\rm SC}$ and $F_{\rm
DW}$, obey the parquet equations shown graphically in Fig.\
\ref{fig:SpinlessSusceptibility} and analytically below:
\begin{mathletters}%
\label{FreeEnergy}
\begin{eqnarray}
   F_{\rm SC}(\xi)&=&\int_0^\xi d\zeta\;{\cal T}_{\rm SC}(\zeta)
   {\cal T}_{\rm SC}^*(\zeta), \\
   F_{\rm DW}(\xi)&=&\int_0^\xi d\zeta\;{\cal T}_{\rm DW}(\zeta)
   {\cal T}_{\rm DW}^*(\zeta).
\end{eqnarray}
\end{mathletters}%
Substituting expressions (\ref{triangular:solutions}) into Eqs.\
(\ref{FreeEnergy}) and dropping the squares of $h_{\rm SC}$ and
$h_{\rm DW}$, we find the susceptibilities:
\begin{mathletters}%
\label{susceptibilities}
\begin{eqnarray}
   \chi_{\rm SC}(\xi)&=&-\bigm[\exp(-2g\xi)-1\bigm]/2g, \\
   \chi_{\rm DW}(\xi)&=&\bigm[\exp(2g\xi)-1\bigm]/2g.
\end{eqnarray}
\end{mathletters}%

   According to Eqs.\ (\ref{susceptibilities}), when the interaction
between electrons is repulsive (attractive), that is, $g$ is positive
(negative), the density-wave (superconducting) susceptibility increases
as temperature decreases ($T\rightarrow0$ and $\xi\rightarrow\infty$):
\begin{equation}
   \chi_{\rm DW(SC)}(\xi)\propto\exp(\pm 2g\xi)
   =\left(\frac{\mu}{ \max\{T,|v_Fq_x|,|\Omega_m|\} } \right)^{\pm2g}.
\label{PowerLaw}
\end{equation}
Susceptibilities (\ref{PowerLaw}) have power dependence on the
temperature and energy, which is one of the characteristic properties
of the Luttinger liquid.  The susceptibilities are finite at finite
temperatures and diverge only at zero temperature, in agreement with
the general theorem \cite{Landau-V} that phase transitions are
impossible at finite temperatures in 1D systems.  Mathematically, the
absence of divergence at finite $\xi$ is due to the cancellation of
the two terms in the r.h.s.\ of Eq.\ (\ref{cancellation}) and
subsequent nonrenormalization of $\gamma(\xi)$.  This nontrivial 1D
result can be obtained only within the parquet, but not the ladder
approximation.

\section{Parquet Equations for One-Dimensional Fermions with Spin}
\label{sec:spin1D}

\begin{figure}
\centerline{\psfig{file=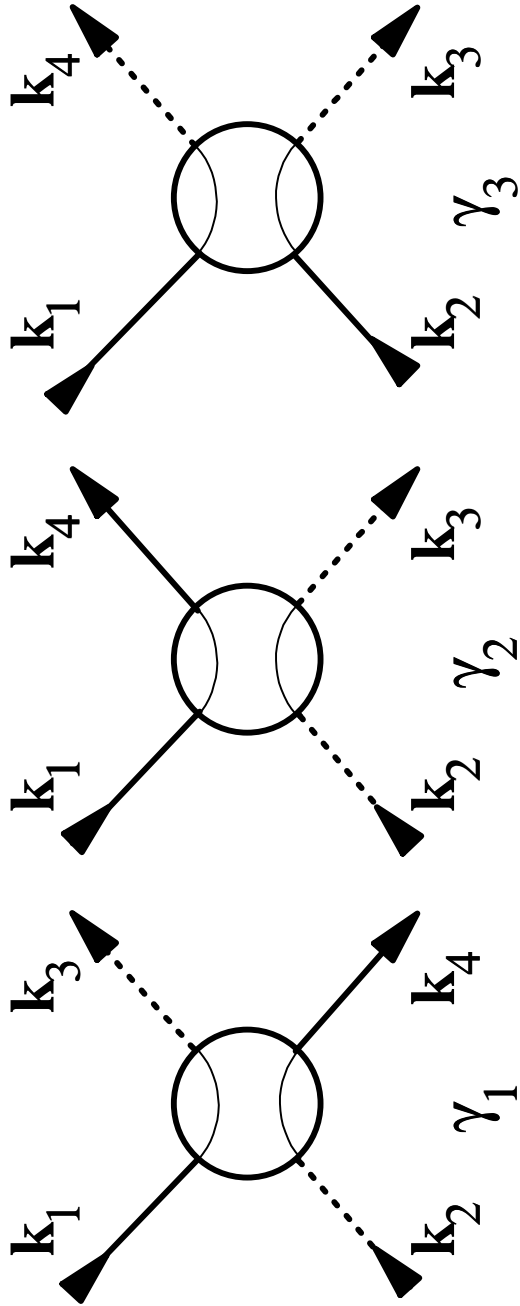,height=0.9\textheight}}
\caption{ Vertices of interaction between electrons with spin:
backward ($\gamma_1$), forward ($\gamma_2$), and umklapp ($\gamma_3$)
scattering.  The thin solid lines inside the circles indicate how spin
is conserved. }
\label{fig:interaction}
\end{figure}

   Now let us consider 1D electrons with spin.  In this case, there
are three vertices of interaction, conventionally denoted as
$\gamma_1$, $\gamma_2$, and $\gamma_3$, which represent backward,
forward, and umklapp scattering, respectively
\cite{Bychkov66,Dzyaloshinskii72a}.  Umklapp scattering should be
considered only when the change of the total momentum of the electrons
in the interaction process, $4k_F$, is equal to the crystal lattice
wave vector, which may or may not be the case in a particular model.
In this paper, we do not consider the vertex $\gamma_4$, which
describes the interaction between the electrons of the same type ($+$
or $-$), because this vertex does not have logarithmic corrections.
The bare Hamiltonian of the interaction, $\hat{H}_{\rm int}$, can be
written as
\begin{eqnarray}
   \hat{H}_{\rm int}&=&\sum_{\sigma,\tau,\rho,\nu=\uparrow\downarrow}
   \int\frac{dk_x^{(1)}}{2\pi}
   \frac{dk_x^{(2)}}{2\pi}\frac{dk_x^{(3)}}{2\pi} 
   \nonumber \\
&& \times\biggm\{
   (-g_1\delta_{\rho\tau}\delta_{\sigma\nu} +
   g_2\delta_{\rho\nu}\delta_{\sigma\tau} )
   \hat{\psi}^+_{\nu+}(k_x^{(1)}+k_x^{(2)}-k_x^{(3)})
   \hat{\psi}^+_{\tau-}(k_x^{(3)})
   \hat{\psi}_{\sigma-}(k_x^{(2)}) \hat{\psi}_{\rho+}(k_x^{(1)})
   \nonumber \\
&& +\left[ g_3\delta_{\rho\nu}\delta_{\sigma\tau}
   \hat{\psi}^+_{\nu-}(k_x^{(1)}+k_x^{(2)}-k_x^{(3)})
   \hat{\psi}^+_{\tau-}(k_x^{(3)})
   \hat{\psi}_{\sigma+}(k_x^{(2)})
   \hat{\psi}_{\rho+}(k_x^{(1)}) + {\rm H.c.} \right]
   \biggm\},
\label{Interaction}
\end{eqnarray}
where the coefficients $g_{1-3}$ denote the bare (unrenormalized)
values of the interaction vertices $\gamma_{1-3}$.  The operators
$\hat{\psi}^+_{\sigma s}$ and $\hat{\psi}_{\sigma s}$ create and
destroy electrons of the type $s=\pm$ and the spin
$\sigma={\uparrow\downarrow}$.  The spin structure of the interaction
Hamiltonian is dictated by conservation of spin.  We picture the
interaction vertices in Fig.\ \ref{fig:interaction}, where the solid
and dashed lines represent the $+$ and $-$ electrons. The thin solid
lines inside the circles indicate how spin is conserved: The spins of
the incoming and outgoing electrons connected by a thin line are the
same.  According to the structure of Hamiltonian (\ref{Interaction}),
the umklapp vertex $\gamma_3$ describes the process where two +
electrons come in and two -- electrons come out, whereas the complex
conjugate vertex $\gamma_3^*$ describes the reversed process.

\begin{figure}
\centerline{\psfig{file=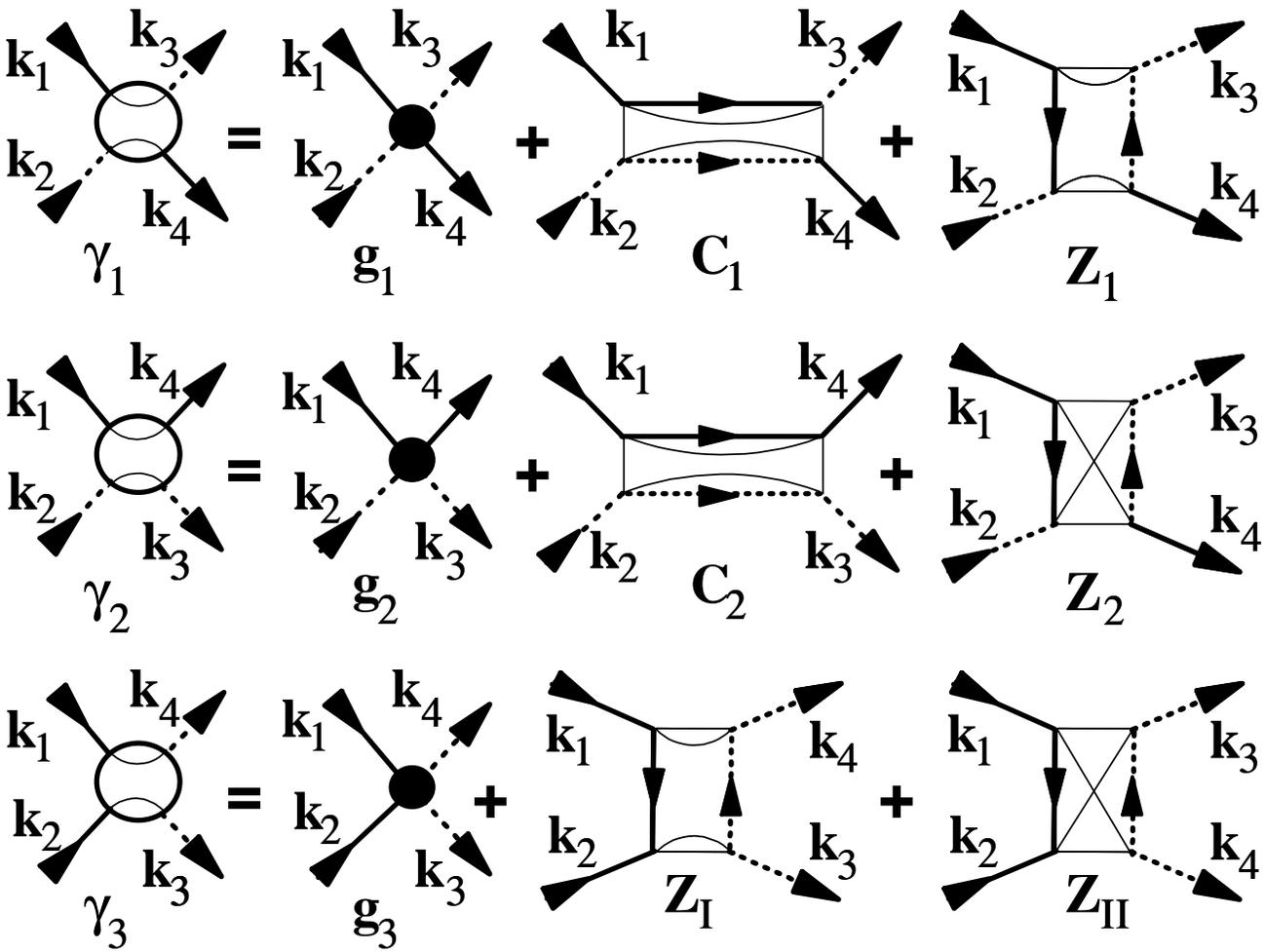,height=0.8\textheight}}
\caption{Decomposition of the interaction vertices (shown as circles)
into superconducting and density-wave bricks (shown as rectangles) for
electrons with spin.  The thin solid lines inside the circles and
rectangles indicate how spin is conserved.  The dots represent the
bare interaction vertices.}
\label{fig:vertices}
\end{figure}

   The three vertices of interaction contain six bricks, as shown
schematically in Fig.\ \ref{fig:vertices}:
\begin{mathletters}%
\label{vertices}
\begin{eqnarray}
   \gamma_1 &=& g_1+C_1+Z_1, \\
   \gamma_2 &=& g_2+C_2+Z_2, \\
   \gamma_3 &=& g_3+Z_I+Z_{II},
\end{eqnarray}
\end{mathletters}%
where $C_1$ and $C_2$ are the superconducting bricks, and $Z_1$,
$Z_2$, $Z_I$, and $Z_{II}$ are the density-wave bricks.  In Fig.\
\ref{fig:vertices}, the thin solid lines inside the bricks represent
spin conservation.  The umklapp vertex has two density-wave bricks
$Z_I$ and $Z_{II}$, which differ in their spin structure.

\begin{figure}
\centerline{\psfig{file=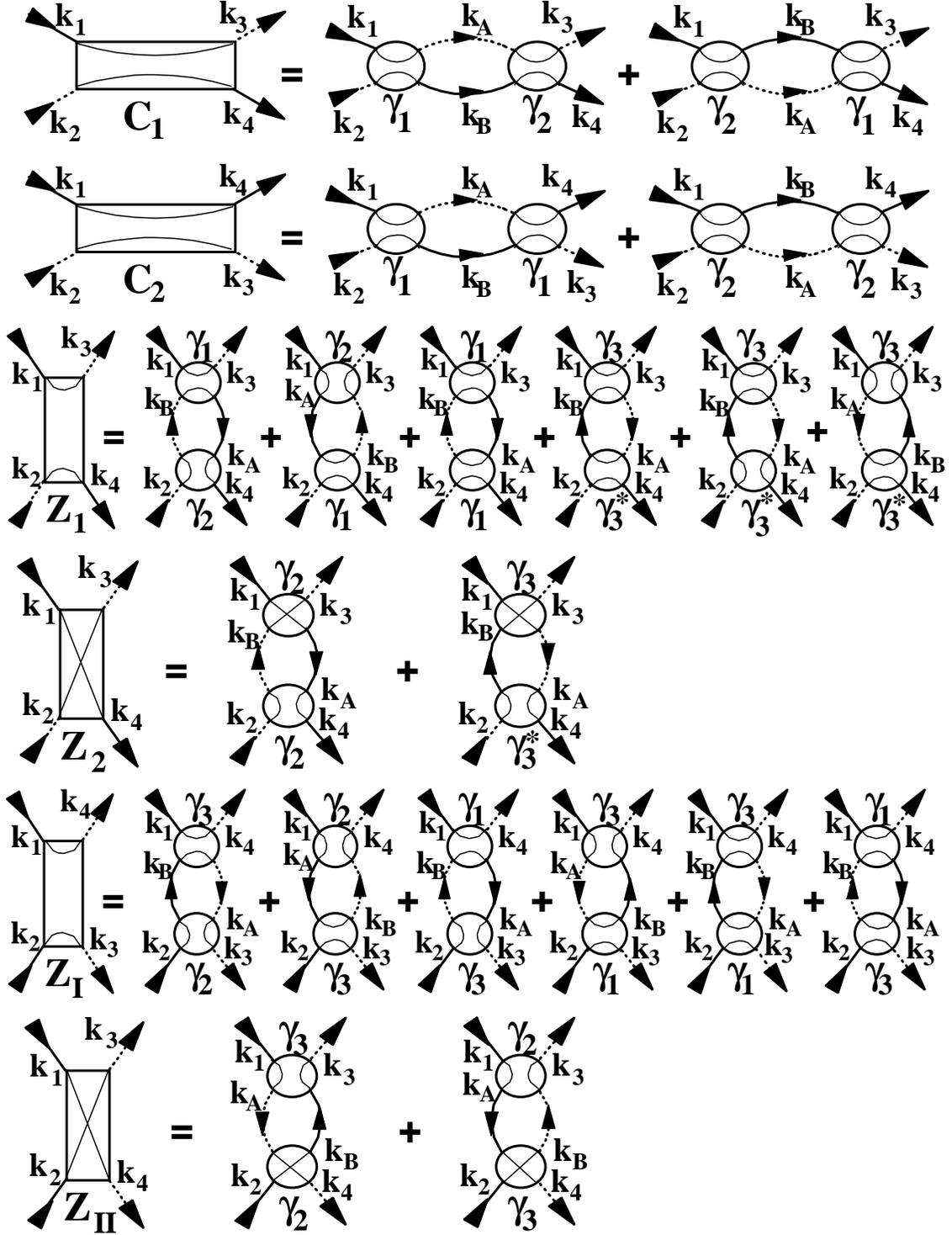,height=0.9\textheight}}
\caption{Parquet equations for the bricks for electrons with spin.
The variables $k_1$ and $k_2$ ($k_3$ and $k_4$) represent momenta of
incoming (outgoing) electrons, whereas the variables $k_{\rm A}$ and
$k_{\rm B}$ represent intermediate momenta that should be integrated
over.}
\label{fig:bricks}
\end{figure}

   Parquet equations for the bricks are derived in the same manner as
in Sec.\ \ref{sec:spinless} by adding appropriate spin structure
dictated by spin conservation.  It is convenient to derive the
equations graphically by demanding that the thin spin lines are
continuous, as shown in Fig.\ \ref{fig:bricks}.  Corresponding
analytic equations can be written using the following rules.  A pair
of parallel (antiparallel) lines connecting two vertices in Fig.\
\ref{fig:bricks} produces the negative (positive) sign.  A closed loop
of the two connecting lines produces an additional factor $-2$ due to
summation over the two spin orientations of the electrons.
\begin{mathletters}%
\label{bricks}
\begin{eqnarray}
   \frac{dC_1(\xi)}{d\xi} &=& -2\gamma_1(\xi)\:\gamma_2(\xi), \\
   \frac{dC_2(\xi)}{d\xi} &=& -\gamma_1^2(\xi)-\gamma_2^2(\xi), \\
   \frac{dZ_1(\xi)}{d\xi} &=& 2\gamma_1(\xi)\:\gamma_2(\xi)
                             -2\gamma_1^2(\xi), \\
   \frac{dZ_2(\xi)}{d\xi} &=& \gamma_2^2(\xi)
                             +\gamma_3(\xi)\gamma_3^*(\xi), \\
   \frac{dZ_I(\xi)}{d\xi} &=& 2\gamma_3(\xi)
                             [\gamma_2(\xi)-\gamma_1(\xi)], \\
   \frac{dZ_{II}(\xi)}{d\xi} &=& 2\gamma_3(\xi)\:\gamma_2(\xi).
\end{eqnarray}
\end{mathletters}%
Combining Eqs.\ (\ref{vertices}) and (\ref{bricks}), we obtain the
well-known closed equations for renormalization of the vertices
\cite{Dzyaloshinskii72a}:
\begin{mathletters}%
\label{RG1D}
\begin{eqnarray}
   \frac{d\gamma_1(\xi)}{d\xi} &=& -2\gamma^2_1(\xi), \\
   \frac{d\gamma_2(\xi)}{d\xi} &=& -\gamma_1^2(\xi)
                                  +\gamma_3(\xi)\gamma_3^*(\xi), \\
   \frac{d\gamma_3(\xi)}{d\xi} &=& 2\gamma_3(\xi)
                                  [2\gamma_2(\xi)-\gamma_1(\xi)].
\end{eqnarray}
\end{mathletters}%

   In the presence of spin, the electron operators in Eq.\
(\ref{Hext}) and, correspondingly, the fields $h_i$ and the triangular
vertices ${\cal T}_i(\xi)$ acquire the spin indices.  Thus, the
superconducting triangular vertex ${\cal T}_{\rm SC}(\xi)$ becomes a
vector:
\begin{equation}
     {\cal T}_{\rm SC}(\xi) = \left( \begin{array}{c}
     {\cal T}_{\rm SC}^{\uparrow \uparrow}(\xi) \\
     {\cal T}_{\rm SC}^{\uparrow \downarrow}(\xi) \\
     {\cal T}_{\rm SC}^{\downarrow \uparrow}(\xi) \\
     {\cal T}_{\rm SC}^{\downarrow \downarrow}(\xi)
     \end{array} \right).
\label{TSC}
\end{equation}
Parquet equations for the triangular vertices are given by the
diagrams shown in Fig.\ \ref{fig:SpinlessTriangle}, where the spin
lines should be added in the same manner as in Fig.\ \ref{fig:bricks}.
The superconducting vertex obeys the following equation:
\begin{equation}
      \frac{d{\cal T}_{\rm SC}(\xi)}{d\xi} =
      \Gamma_{\rm SC}(\xi)\;{\cal T}_{\rm SC}(\xi),
\label{MTRX}
\end{equation}
where the matrix $\Gamma_{\rm SC}(\xi)$ is
\begin{equation}
     \Gamma_{\rm SC}(\xi) = \left( \begin{array}{cccc}
     -\gamma_2 + \gamma_1 & 0 & 0 &0 \\
     0 & -\gamma_2 & \gamma_1 & 0  \\
     0 & \gamma_1 & -\gamma_2 & 0 \\
     0 & 0 & 0 & -\gamma_2 + \gamma_1  \end{array} \right).
\label{GSC}
\end{equation}
Linear equation (\ref{MTRX}) is diagonalized by introducing the
singlet, ${\cal T}_{\rm SSC}$, and the triplet, ${\cal T}_{\rm TSC}$,
superconducting triangular vertices:
\begin{mathletters}%
\label{SC}
\begin{eqnarray}
  {\cal T}_{\rm SSC}(\xi) &=&
     {\cal T}_{\rm SC}^{\uparrow \downarrow}(\xi) -
     {\cal T}_{\rm SC}^{\downarrow \uparrow}(\xi) ,
\label{SCS}  \\
  {\cal T}_{\rm TSC}(\xi) &=& \left( \begin{array}{c}
     {\cal T}_{\rm SC}^{\uparrow \uparrow}(\xi) \\
     {\cal T}_{\rm SC}^{\uparrow \downarrow}(\xi) +
     {\cal T}_{\rm SC}^{\downarrow \uparrow}(\xi) \\
     {\cal T}_{\rm SC}^{\downarrow \downarrow}(\xi)
     \end{array} \right),
\label{SCT}
\end{eqnarray}
\end{mathletters}%
which obey the following equations:
\begin{equation}
      \frac{d{\cal T}_{\rm SSC(TSC)}(\xi)}{d\xi} =
      [\mp\gamma_1(\xi)-\gamma_2(\xi)]
      \;{\cal T}_{\rm SSC(TSC)}(\xi).
\label{SCOP}
\end{equation}
In Eq.\ (\ref{SCOP}) the sign $-$ and the index SSC correspond to the
singlet superconductivity, whereas the sign $+$ and the index TSC
correspond to the triplet one.  In the rest of the paper, we use the
index SC where discussion applies to both SSC and TSC.

   Now let us consider the density-wave triangular vertices, first in
the absence of umklapp.  They form a vector
\begin{equation}
     {\cal T}_{\rm DW}(\xi) = \left( \begin{array}{c}
     {\cal T}_{\rm DW}^{\uparrow \uparrow}(\xi) \\
     {\cal T}_{\rm DW}^{\uparrow \downarrow}(\xi) \\
     {\cal T}_{\rm DW}^{\downarrow \uparrow}(\xi) \\
     {\cal T}_{\rm DW}^{\downarrow \downarrow}(\xi)
     \end{array} \right),
\label{TDW}
\end{equation}
which obeys the equation
\begin{equation}
      \frac{d{\cal T}_{\rm DW}(\xi)}{d\xi} =
      \Gamma_{\rm DW}(\xi)\;{\cal T}_{\rm DW}(\xi)
\label{DWMTRX}
\end{equation}
with the matrix
\begin{equation}
     \Gamma_{\rm DW}(\xi) = \left( \begin{array}{cccc}
     -\gamma_1 + \gamma_2 & 0 & 0 &-\gamma_1 \\
     0 & \gamma_2 & 0 & 0  \\
     0 & 0 & \gamma_2 & 0 \\
     -\gamma_1 & 0 & 0 & -\gamma_1 + \gamma_2  \end{array} \right).
\label{GDW}
\end{equation}

   Eq.\ (\ref{DWMTRX}) is diagonalized by introducing the charge-,
${\cal T}_{\rm CDW}$, and the spin-, ${\cal T}_{\rm SDW}$, density-wave
triangular vertices:
\begin{mathletters}%
\label{DW}
\begin{eqnarray}
  {\cal T}_{\rm CDW}(\xi) &=&
     {\cal T}_{\rm DW}^{\uparrow \uparrow}(\xi) +
     {\cal T}_{\rm DW}^{\downarrow \downarrow}(\xi) ,
\label{DWS}  \\
  {\cal T}_{\rm SDW}(\xi) &=& \left( \begin{array}{c}
      {\cal T}_{\rm DW}^{\uparrow \downarrow}(\xi) \\
      {\cal T}_{\rm DW}^{\downarrow \uparrow}(\xi) \\
      {\cal T}_{\rm DW}^{\uparrow \uparrow}(\xi) -
      {\cal T}_{\rm DW}^{\downarrow \downarrow}(\xi)
      \end{array} \right),
\label{DWT}
\end{eqnarray}
\end{mathletters}%
which obey the following equations:
\begin{mathletters}%
\label{DWOP}
\begin{eqnarray}
   \frac{d{\cal T}_{\rm CDW}(\xi)}{d\xi} &=&
      [-2\gamma_1(\xi)+\gamma_2(\xi)]\;{\cal T}_{\rm CDW}(\xi), \\
   \frac{d{\cal T}_{\rm SDW}(\xi)}{d\xi} &=&
      \gamma_2(\xi)\;{\cal T}_{\rm SDW}(\xi).
\end{eqnarray}
\end{mathletters}%

   When the umklapp vertices $\gamma_3$ and $\gamma_3^*$ are introduced,
they become offdiagonal matrix elements in Eqs.\ (\ref{DWOP}), mixing
${\cal T}_{\rm CDW}$ and ${\cal T}_{\rm SDW}$ with their complex
conjugates.  Assuming for simplicity that $\gamma_3$ is real, we find
that the following linear combinations diagonalize the equations:
\begin{equation}
   {\cal T}_{{\rm CDW(SDW)}\pm}={\cal T}_{\rm CDW(SDW)}
                                \pm {\cal T}^*_{\rm CDW(SDW)},
\label{DW+-}
\end{equation}
and the equations become:
\begin{mathletters}%
\label{DWOP+-}
\begin{eqnarray}
   \frac{d{\cal T}_{{\rm CDW}\pm}(\xi)}{d\xi} &=&
      [-2\gamma_1(\xi)+\gamma_2(\xi)\mp\gamma_3(\xi)]
      \;{\cal T}_{{\rm CDW}\pm}(\xi), \\
   \frac{d{\cal T}_{{\rm SDW}\pm}(\xi)}{d\xi} &=&
      [\gamma_2(\xi)\pm\gamma_3(\xi)]\;{\cal T}_{{\rm SDW}\pm}(\xi).
\end{eqnarray}
\end{mathletters}%

   If the external fields $h_i$ are set to unity in the initial
conditions of the type (\ref{triangular}) for all triangular vertices
$i$ = SSC, TSC, CDW$\pm$, and SDW$\pm$, then the corresponding
susceptibilities are equal numerically to the free energy corrections of
the type (\ref{FreeEnergy}):
\begin{equation}
   \chi_i(\xi)= \int_0^\xi d\zeta\;
   {\cal T}_i(\zeta){\cal T}_i^*(\zeta).
\label{chii}
\end{equation}

   Eqs.\ (\ref{RG1D}), (\ref{SCOP}), (\ref{DWOP+-}), and (\ref{chii})
were solved analytically in Ref.\ \cite{Dzyaloshinskii72a}, where a
complete phase diagram of the 1D electron gas with spin was obtained.

\section{Parquet Equations for Two-Dimensional Electrons}
\label{sec:2D}

\begin{figure}
\centerline{\psfig{file=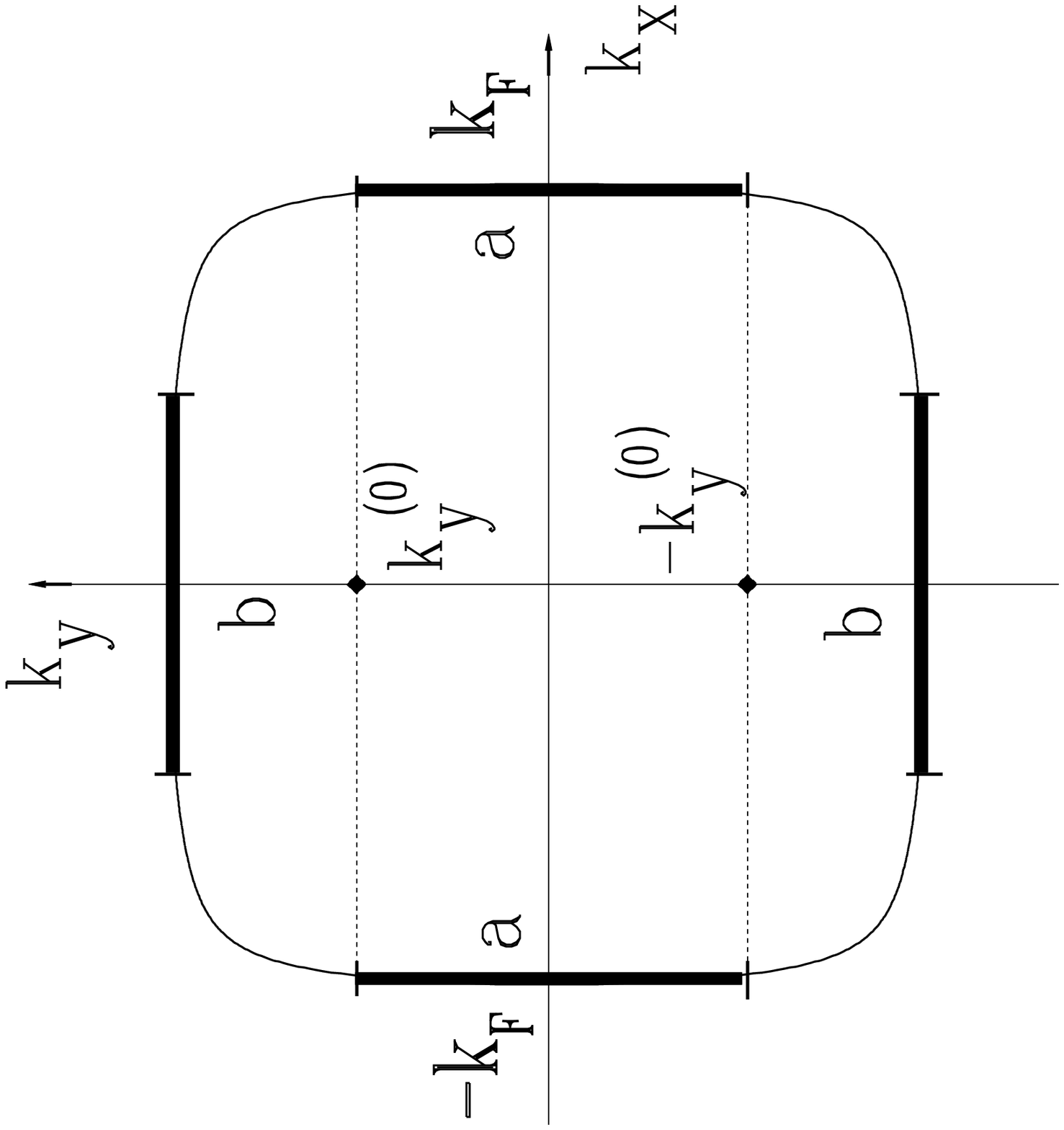,height=0.9\textheight}}
\caption{Fermi surface of a 2D electron gas.  The thick lines
indicate flat regions on the Fermi surface.}
\label{fig:2DFS}
\end{figure}

   Now let us consider a 2D electron gas with the Fermi surface shown
schematically in Fig.\ \ref{fig:2DFS}.  It contains two pairs of flat
regions, shown as the thick lines and labeled by the letters $a$ and
$b$.  Such a Fermi surface resembles the Fermi surfaces of some
high-$T_c$ superconductors \cite{ZXShen93}.  In our consideration, we
restrict the momenta of electrons to the flat sections only.  In this
way, we effectively neglect the rounded portions of the Fermi surface,
which are not relevant for the parquet consideration, because the
density-wave loop is not divergent there.  One can check also that the
contributions of the portions $a$ and $b$ do not mix with each other
in the parquet manner, so they may be treated separately. For this
reason, we will consider only the region $a$, where the 2D electron
states are labeled by the two momenta $k_x$ and $k_y$, the latter
momentum being restricted to the interval $[-k_y^{(0)},k_y^{(0)}]$.
In our model, the energy of electrons depends only on the momentum
$k_x$ according to Eq.\ (\ref{eps}) and does not depend on the
momentum $k_y$.  We neglect possible dependence of the Fermi velocity
$v_F$ on $k_y$; it was argued in Ref.\ \cite{Luther94} that this
dependence is irrelevant in the renormalization-group sense.

   In the 2D case, each brick or vertex of interaction between
electrons acquires extra variables $k_y^{(1)}$, $k_y^{(2)}$, and
$k_y^{(3)}$ in addition to the 1D variables
$\omega_1,\:\omega_2,\:\omega_3,\:v_Fk_x^{(1)},\:v_Fk_x^{(2)}$, and
$v_Fk_x^{(3)}$.  These two sets of variables play very different
roles.  The Green functions, which connect the vertices and produce
the logarithms $\xi$, depend only on the second set of variables.
Thus, following the parquet approach outlined in the previous
Sections, we dump all the $\omega$ and $v_Fk_x$ variables of a vertex
or a brick into a single variable $\xi$.  At the same time, the
$k_y^{(1)}$, $k_y^{(2)}$, and $k_y^{(3)}$ variables remain independent
and play the role of indices labeling the vertices, somewhat similar
to the spin indices.  Thus, each vertex and brick is a function of
several variables, which we will always write in the following order:
$\gamma(k_y^{(1)},k_y^{(2)};\:k_y^{(3)},k_y^{(4)};\:\xi)$.  It is
implied that the first four variables satisfy the momentum
conservation law $k_y^{(1)}+k_y^{(2)}=k_y^{(3)}+k_y^{(4)}$, and each
of them belongs to the interval $[-k_y^{(0)},k_y^{(0)}]$.  The
assignment of the variables $k_y^{(1)}$, $k_y^{(2)}$, $k_y^{(3)}$, and
$k_y^{(4)}$ to the ends of the vertices and bricks is shown in Fig.\
\ref{fig:vertices}, where the labels $k_j$ ($j=1-4$) should be
considered now as the variables $k_y^{(j)}$.  To shorten notation, it
is convenient to combine these variable into a single four-component
vector
\begin{equation}
   {\cal K}=(k_y^{(1)},k_y^{(2)};\:k_y^{(3)},k_y^{(4)}),
\label{K}
\end{equation}
so that the relation between the vertices and the bricks can be
written as
\begin{mathletters}%
\label{2Dgammas}
\begin{eqnarray}
   \gamma_1({\cal K},\xi) &=&
      g_1 + C_1({\cal K},\xi) + Z_1({\cal K},\xi),\\
   \gamma_2({\cal K},\xi) &=&
      g_2 + C_2({\cal K},\xi) + Z_2({\cal K},\xi),\\
   \gamma_3({\cal K},\xi) &=&
      g_3 + Z_I({\cal K},\xi) + Z_{II}({\cal K},\xi).
\end{eqnarray}
\end{mathletters}%

   After this introduction, we are in a position to write the parquet
equations for the bricks.  These equations are shown graphically in
Fig.\ \ref{fig:bricks}, where again the momenta $k_j$ should be
understood as $k_y^{(j)}$.  Analytically, the equations are written
below, with the terms in the same order as in Fig.\ \ref{fig:bricks}:
\begin{mathletters}%
\label{2Dbricks}
\begin{eqnarray}
      \frac{\partial C_1({\cal K},\xi)}{\partial \xi} &=&
      -\gamma_1({\cal K}_1,\xi)\circ\gamma_2({\cal K}_1^{\prime},\xi) -
      \gamma_2({\cal K}_1,\xi)\circ\gamma_1({\cal K}_1^{\prime},\xi),
\label{C1} \\
      \frac{\partial C_2({\cal K},\xi)}{\partial \xi} &=&
      -\gamma_1({\cal K}_1,\xi)\circ\gamma_1({\cal K}_1^{\prime},\xi)
      -\gamma_2({\cal K}_1,\xi)\circ\gamma_2({\cal K}_1^{\prime},\xi),
\label{C2}  \\
      \frac{\partial Z_1({\cal K},\xi)}{\partial \xi} &=&
      \gamma_1({\cal K}_2,\xi)\circ\gamma_2({\cal K}_2^{\prime},\xi) +
      \gamma_2({\cal K}_2,\xi)\circ\gamma_1({\cal K}_2^{\prime},\xi)
      - 2 \gamma_1({\cal K}_2,\xi)\circ\gamma_1({\cal K}_2^{\prime},\xi)
\nonumber \\
   && - 2 \tilde{\gamma}_3({\cal K}_2,\xi)
      \circ\tilde{\bar{\gamma}}_3({\cal K}_2^{\prime},\xi)
      + \tilde{\gamma}_3({\cal K}_2,\xi)
      \circ\bar{\gamma}_3({\cal K}_2^{\prime},\xi)
      + \gamma_3({\cal K}_2,\xi)
      \circ\tilde{\bar{\gamma}}_3({\cal K}_2^{\prime},\xi),
\label{Z1}  \\
      \frac{\partial Z_2({\cal K},\xi)}{\partial \xi} &=&
      \gamma_2({\cal K}_2,\xi)\circ\gamma_2({\cal K}_2^{\prime},\xi)
     +\gamma_3({\cal K}_2,\xi)\circ\bar{\gamma}_3({\cal K}_2^{\prime},\xi),
\label{Z2} \\
      \frac{\partial Z_I({\cal K},\xi)}{\partial \xi} &=&
      \tilde{\gamma}_3({\cal K}_3,\xi)
      \circ\gamma_2({\cal K}_3^{\prime},\xi)
      + \gamma_2({\cal K}_3,\xi)
      \circ\tilde{\gamma}_3({\cal K}_3^{\prime},\xi) +
      \gamma_1({\cal K}_3,\xi)\circ\gamma_3({\cal K}_3^{\prime},\xi)
\nonumber \\
  &&  + \gamma_3({\cal K}_3,\xi)\circ\gamma_1({\cal K}_3^{\prime},\xi)
      - 2\tilde{\gamma}_3({\cal K}_3,\xi)
       \circ\gamma_1({\cal K}_3^{\prime},\xi) -
       2\gamma_1({\cal K}_3,\xi)
       \circ\tilde{\gamma}_3({\cal K}_3^{\prime},\xi),
\label{ZI} \\
      \frac{\partial Z_{II}({\cal K},\xi)}{\partial \xi} &=&
      \gamma_3({\cal K}_2,\xi)\circ\gamma_2({\cal K}_2'',\xi)
      + \gamma_2({\cal K}_2,\xi)\circ\gamma_3({\cal K}_2'',\xi),
\label{ZII}
\end{eqnarray}
\end{mathletters}%
where
\begin{mathletters}%
\label{2DK}
\begin{eqnarray}
   && {\cal K}_1=(k_y^{(1)},k_y^{(2)};\:k_y^{(A)},k_y^{(B)}),\quad
      {\cal K}_1'=(k_y^{(B)},k_y^{(A)};\:k_y^{(3)},k_y^{(4)}),\\
   && {\cal K}_2=(k_y^{(1)},k_y^{(B)};\:k_y^{(3)},k_y^{(A)}),\quad
      {\cal K}_2'=(k_y^{(A)},k_y^{(2)};\:k_y^{(B)},k_y^{(4)}),\quad
      {\cal K}_2''=(k_y^{(2)},k_y^{(A)};\:k_y^{(4)},k_y^{(B)}),\\
   && {\cal K}_3=(k_y^{(1)},k_y^{(B)};\:k_y^{(4)},k_y^{(A)}),\quad
      {\cal K}_3'=(k_y^{(2)},k_y^{(A)};\:k_y^{(3)},k_y^{(B)}),
\end{eqnarray}
\end{mathletters}%
and the tilde and the bar operations are defined as
\begin{mathletters}%
\label{TildeBar}
\begin{eqnarray}
   \tilde{\gamma}_j(k_y^{(1)},k_y^{(2)};\:k_y^{(3)},k_y^{(4)};\:\xi)
   &\equiv&\gamma_j(k_y^{(1)},k_y^{(2)};\:k_y^{(4)},k_y^{(3)};\:\xi),
\label{tilde}\\
   \bar{\gamma}_3(k_y^{(1)},k_y^{(2)};\:k_y^{(3)},k_y^{(4)};\:\xi)
   &\equiv&\gamma_3^*(k_y^{(4)},k_y^{(3)};\:k_y^{(2)},k_y^{(1)};\:\xi).
\end{eqnarray}
\end{mathletters}%
In Eqs.\ (\ref{2Dbricks}), we introduced the operation $\circ$ that
represents the integration over the internal momenta of the loops in
Fig.\ \ref{fig:bricks}.  It denotes the integration over the
intermediate momentum $k_y^{(A)}$ with the restriction that both
$k_y^{(A)}$ and $k_y^{(B)}$, another intermediate momentum determined
by conservation of momentum, belong to the interval
$[-k_y^{(0)},k_y^{(0)}]$.  For example, the explicit form of the first
term in the r.h.s.\ of Eq.\ (\ref{C1}) is:
\begin{eqnarray}
   &&\gamma_1({\cal K}_1,\xi)\circ\gamma_2({\cal K}_1^{\prime},\xi)=
   \displaystyle
   \int_{ -k_y^{(0)} \leq k_y^{(A)} \leq k_y^{(0)};\;\;
       -k_y^{(0)} \leq k_y^{(1)}+k_y^{(2)}-k_y^{(A)} \leq k_y^{(0)} }
   \frac{\displaystyle dk_y^{(A)}}{\displaystyle 2\pi}\,
\nonumber \\ && \times
\gamma_1(k_y^{(1)},k_y^{(2)};\:k_y^{(A)},k_y^{(1)}+k_y^{(2)}-k_y^{(A)};\:\xi)
\,
\gamma_2(k_y^{(1)}+k_y^{(2)}-k_y^{(A)},k_y^{(A)};\:k_y^{(3)},k_y^{(4)};\:\xi).
\label{o}
\end{eqnarray}
Eqs.\ (\ref{2Dbricks}) and (\ref{2Dgammas}) with definitions
(\ref{K}), (\ref{2DK}), and (\ref{TildeBar}) form a closed system of
integrodifferential equations, which will be solved numerically in
Sec.\ \ref{sec:numerical}. The initial conditions for Eqs.\
(\ref{2Dbricks}) and (\ref{2Dgammas}) are that all the $C$ and $Z$
bricks are equal to zero at $\xi=0$.

\begin{figure}
\centerline{\psfig{file=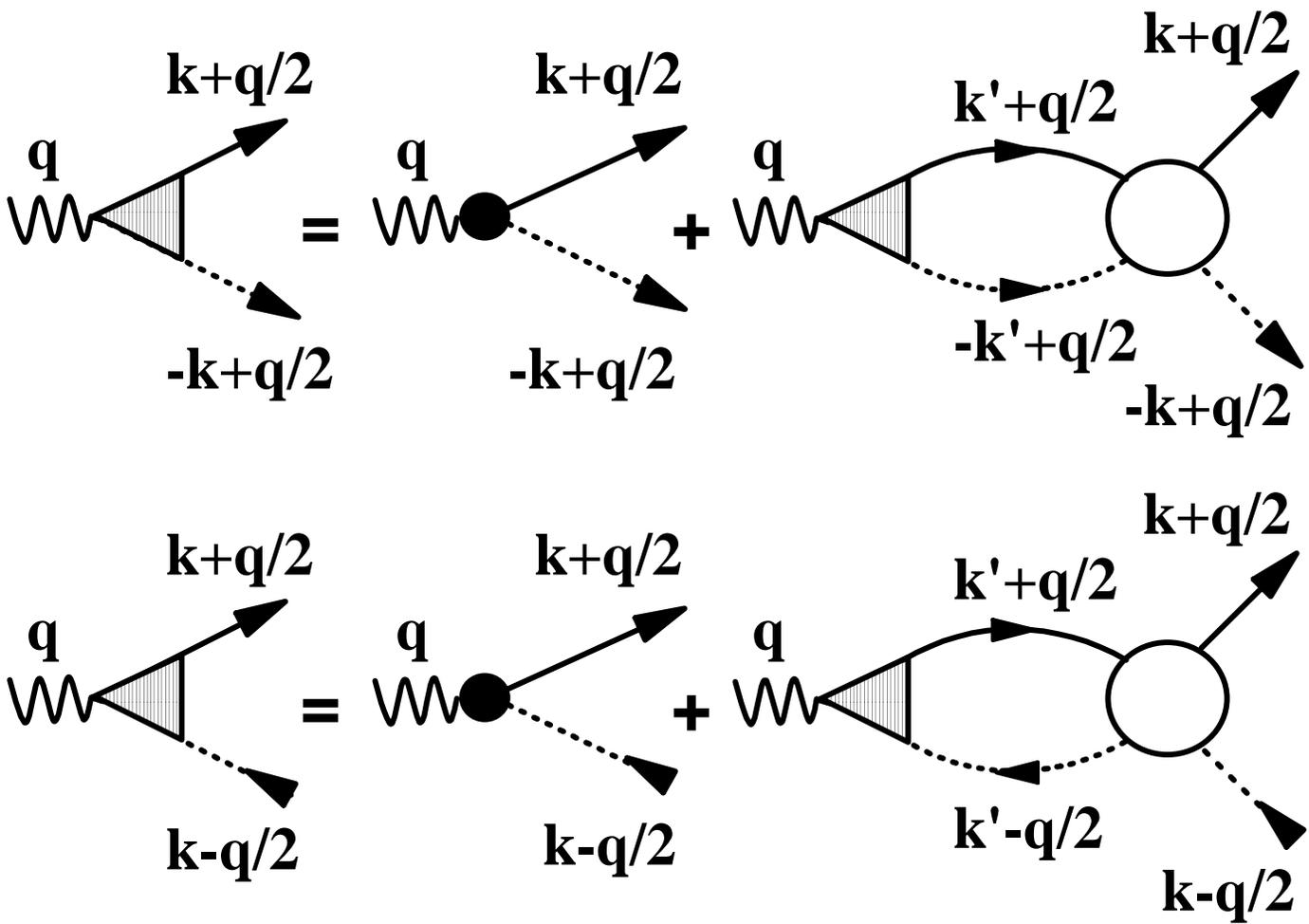,height=0.9\textheight}}
\caption{Parquet equations for the triangular vertices in the
spinless case.  The filled triangles represent the vertices ${\cal
T}_{\rm SC}$ and ${\cal T}_{\rm DW}$, whereas the dots represent the
auxiliary external fields $h_{\rm SC}$ and $h_{\rm DW}$.}
\label{fig:SpinlessTriangle}
\end{figure}

   Parquet equations for the superconducting triangular vertices can
be found in the 2D case by adding the $k_y$ momenta to the 1D
equations (\ref{SCOP}).  The equations are shown graphically in Fig.\
\ref{fig:SpinlessTriangle}, where the momenta $k$ and $q$ should be
interpreted as $k_y$ and $q_y$:
\begin{equation}
   \frac{\partial{\cal T}_{\rm SSC(TSC)}(k_y,q_y,\xi)}{\partial\xi}
   = f_{\rm SSC(TSC)}({\cal K}_{\rm SC},\xi)\circ
   {\cal T}_{\rm SSC(TSC)}(k'_y,q_y,\xi),
\label{2DSCOP}
\end{equation}
where
\begin{eqnarray}
   && f_{\rm SSC(TSC)}({\cal K}_{\rm SC},\xi)=
      \mp\gamma_1({\cal K}_{\rm SC},\xi)-
      \gamma_2({\cal K}_{\rm SC},\xi),
\label{fSC} \\
   && {\cal K}_{\rm SC}=(k'_y+q_y/2,-k'_y+q_y/2; -k_y+q_y/2,k_y+q_y/2),
\end{eqnarray}
and the operator $\circ$ denotes the integration over $k'_y$ with the
restriction that both $k'_y+q_y/2$ and $-k'_y+q_y/2$ belong to the
interval $[-k_y^{(0)},k_y^{(0)}]$.  The $\mp$ signs in front of
$\gamma_1$ in Eq.\ (\ref{fSC}) correspond to the singlet and triplet
superconductivity.  As discussed in Sec.\ \ref{sec:spinless}, the
triangular vertex ${\cal T}_{\rm SC}(k_y,q_y,\xi)$ is the
superconducting order parameter, $q_y$ and $k_y$ being the
$y$-components of the total and the relative momenta of the electrons
in a Cooper pair.  Indeed, the vertex ${\cal T}_{\rm SC}(k_y,q_y,\xi)$
obeys the linear equation shown in Fig.\ \ref{fig:SpinlessTriangle},
which is the linearized Gorkov equation for the superconducting order
parameter.  As the system approaches a phase transition, the vertex
${\cal T}_{\rm SC}(k_y,q_y,\xi)$ diverges in overall magnitude, but
its dependence on $k_y$ for a fixed $q_y$ remains the same, up to a
singular, $\xi$-dependent factor.  The dependence of ${\cal T}_{\rm
SC}(k_y,q_y,\xi)$ on $k_y$ describes the distribution of the emerging
order parameter over the Fermi surface.  The numerically found
behavior of ${\cal T}_{\rm SC}(k_y,q_y,\xi)$ is discussed in Sec.\
\ref{sec:numerical}.

   Due to the particular shape of the Fermi surface, the vertices of
interaction in our 2D model have two special symmetries: with respect to
the sign change of all momenta $k_y$ and with respect to the exchange of
the $+$ and $-$ electrons:
\begin{mathletters}%
\label{symmetry}
\begin{eqnarray}
   \gamma_i(k_y^{(1)},k_y^{(2)};\:k_y^{(3)},k_y^{(4)};\:\xi) &=&
   \gamma_i(-k_y^{(1)},-k_y^{(2)};\:-k_y^{(3)},-k_y^{(4)};\:\xi),
   \quad i=1,2,3; \\
   \gamma_i(k_y^{(1)},k_y^{(2)};\:k_y^{(3)},k_y^{(4)};\:\xi) &=&
   \gamma_i(k_y^{(2)},k_y^{(1)};\:k_y^{(4)},k_y^{(3)};\:\xi),
   \quad i=1,2,3; \\
   \gamma_3(k_y^{(1)},k_y^{(2)};\:k_y^{(3)},k_y^{(4)};\:\xi) &=&
   \gamma_3(k_y^{(4)},k_y^{(3)};\:k_y^{(2)},k_y^{(1)};\:\xi),
   \label{*}
\end{eqnarray}
\end{mathletters}%
where in Eq.\ (\ref{*}) we assume for simplicity that $\gamma_3$ is
real.  As a consequence of (\ref{symmetry}), Eqs.\ (\ref{2DSCOP}) are
invariant with respect to the sign reversal of $k_y$ in ${\cal T}_{\rm
SC}(k_y,q_y,\xi)$ at a fixed $q_y$.  The following combinations of the
triangular vertices form two irreducible representations of this
symmetry, that is, they are independent and do not mix in Eqs.\
(\ref{2DSCOP}):
\begin{equation}
   {\cal T}^\pm_{\rm SSC(TSC)}(k_y,q_y,\xi)=
   {\cal T}_{\rm SSC(TSC)}(k_y,q_y,\xi)
   \pm {\cal T}_{\rm SSC(TSC)}(-k_y,q_y,\xi).
\label{SASC}
\end{equation}
The triangular vertices ${\cal T}^\pm_{\rm SSC(TSC)}(k_y,q_y,\xi)$
describe the superconducting order parameters that are either
symmetric or antisymmetric with respect to the sign change of $k_y$.
When ${\cal T}^+_{\rm SSC}$ is extended over the whole 2D Fermi
surface (see Fig.\ \ref{fig:2DFS}), it acquires the $s$-wave symmetry,
whereas ${\cal T}^-_{\rm SSC}$ the $d$-wave symmetry.  The symmetrized
vertices ${\cal T}^\pm_{\rm SSC(TSC)}(k_y,q_y,\xi)$ obey the same
Eqs.\ (\ref{2DSCOP}) as the unsymmetrized ones.

   The equations for the density-wave triangular vertices are obtained
in a similar manner:
\begin{mathletters}%
\label{CSDWOPA}
\begin{eqnarray}
   \frac{\partial{\cal T}_{{\rm CDW}\pm}^{\pm}(k_y,q_y,\xi)}
      {\partial\xi} &=&
   f_{{\rm CDW}\pm}({\cal K}_{\rm DW},\xi)\circ
      {\cal T}_{{\rm CDW}\pm}^{\pm}(k'_y,q_y,\xi),
\label{CDWOPA} \\
   \frac{\partial{\cal T}_{{\rm SDW}\pm}^{\pm}(k_y,q_y,\xi)}
      {\partial\xi} &=&
   f_{{\rm SDW}\pm}({\cal K}_{\rm DW},\xi)\circ
      {\cal T}_{{\rm SDW}\pm}^{\pm}(k'_y,q_y,\xi),
\label{SDWOPA}
\end{eqnarray}
\end{mathletters}%
where
\begin{eqnarray}
   && f_{{\rm  CDW}\pm}({\cal K}_{\rm DW},\xi)=
      -2\gamma_1({\cal K}_{\rm DW},\xi)
      \mp 2\tilde{\gamma}_3({\cal K}_{\rm DW},\xi)
      +\gamma_2({\cal K}_{\rm DW},\xi)
      \pm \gamma_3({\cal K}_{\rm DW},\xi),
\label{fCDW} \\
   && f_{{\rm  SDW}\pm}({\cal K}_{\rm DW},\xi)=
      \gamma_2({\cal K}_{\rm DW},\xi) \pm \gamma_3({\cal K}_{\rm DW},\xi),
\label{fSDW} \\
   && {\cal K}_{\rm DW} = (k'_y+q_y/2,k_y-q_y/2; k'_y-q_y/2,k_y+q_y/2).
\end{eqnarray}
The $\pm$ signs in the subscripts of ${\cal T}$ in Eqs.\
(\ref{CSDWOPA}) and in front of $\gamma_3$ in Eqs.\
(\ref{fCDW})--(\ref{fSDW}) refer to the umklapp symmetry discussed in
Sec.\ \ref {sec:spin1D}, whereas the $\pm$ signs in the superscripts
of ${\cal T}$ refer to the symmetry with respect to sign reversal of
$k_y$, discussed above in the superconducting case.  The
$k_y$-antisymmetric density waves are actually the waves of charge
current and spin current \cite{Halperin68,Dzyaloshinskii87a}, also
known in the so-called flux phases \cite{FluxPhases}.

   Once the triangular vertices ${\cal T}_i$ are found, the
corresponding susceptibilities $\chi_i$ are calculated according to
the following equation, similar to Eq.\ (\ref{chii}):
\begin{equation}
   \chi_i(q_y,\xi)=\int_0^\xi d\zeta \int\frac{dk_y}{2\pi}
   {\cal T}_i(k_y,q_y,\zeta){\cal T}_i^*(k_y,q_y,\zeta),
\label{2Dchii}
\end{equation}
where the integration over $k_y$ is restricted so that both
$k_y\pm q_y/2$ belong to the interval $[-k_y^{(0)},k_y^{(0)}]$.

   Using functions (\ref{fSC}), (\ref{fCDW}), and (\ref{fSDW}) and
symmetries (\ref{symmetry}), we can rewrite Eqs.\ (\ref{2Dbricks}) in
a more compact form.  For that purpose, we introduce the SSC, TSC,
CDW, and SDW bricks that are the linear combinations of the original
bricks:
\begin{mathletters}%
\label{NEWbricks}
\begin{eqnarray}
   C_{\rm SSC(TSC)} &=& C_2 \pm C_1,
\label{CST} \\
   Z_{{\rm CDW}\pm} &=& \tilde{Z}_2 - 2 \tilde{Z}_1
   \pm (\tilde{Z}_{II} - 2 Z_I),
\label{ZCW} \\
   Z_{{\rm SDW}\pm} &=& Z_2 \pm Z_{II},
\label{ZSW}
\end{eqnarray}
\end{mathletters}%
where the tilde operation is defined in Eq.\ (\ref{tilde}).  Then,
Eqs.\ (\ref{2Dbricks}) become:
\begin{mathletters}%
\label{NEWRG}
\begin{eqnarray}
    \frac{\partial C_{\rm SSC(TSC)}({\cal K},\xi)}{\partial \xi}
    &=& -f_{\rm SSC(TSC)}({\cal K}_1,\xi)
    \circ f_{\rm SSC(TSC)}({\cal K}_1^{\prime},\xi),
\label{SC1} \\
    \frac{\partial Z_{{\rm CDW}\pm}({\cal K},\xi)}{\partial \xi}
    &=& f_{{\rm CDW}\pm}({\cal K}_3,\xi)
    \circ f_{{\rm CDW}\pm}({\cal K}_3^{\prime},\xi),
\label{CW1} \\
    \frac{\partial Z_{{\rm SDW}\pm}({\cal K},\xi)}{\partial \xi}
    &=& f_{{\rm SDW}\pm}({\cal K}_2,\xi)
    \circ f_{{\rm SDW}\pm}({\cal K}_2^{\prime},\xi).
\label{SW1}
\end{eqnarray}
\end{mathletters}%
The parquet equations in the form (\ref{NEWRG}) were obtained in
Ref.\ \cite{Dzyaloshinskii72b}.

   It is instructive to trace the difference between the parquet
equations (\ref{NEWRG}) and the corresponding ladder equations.  Suppose
that, for some reason, only one brick, say $C_{\rm SSC}$, among the six
bricks (\ref{NEWbricks}) is appreciable, whereas the other bricks may be
neglected.  Using definitions (\ref{2Dgammas}) and (\ref{fSC}), we find
that Eq.\ (\ref{SC1}) becomes a closed equation:
\begin{equation}
    \frac{\partial f_{\rm SSC}({\cal K},\xi)}{\partial \xi} =
    f_{\rm SSC}({\cal K}_1,\xi)\circ f_{\rm SSC}({\cal K}_1',\xi),
\label{fSCRG}
\end{equation}
where
\begin{equation}
    f_{\rm SSC}({\cal K}_1,\xi)=-g_1-g_2-C_{\rm SSC}({\cal K},\xi).
\label{fSSC}
\end{equation}
Eq.\ (\ref{fSCRG}) is the ladder equation for the singlet
superconductivity.  When the initial value $-(g_1+g_2)$ of the vertex
$f_{\rm SSC}$ is positive, Eq.\ (\ref{fSCRG}) has a singular solution
($f_{\rm SSC}\rightarrow\infty$ at $\xi\rightarrow\xi_c$), which
describes a phase transition into the singlet superconducting state at
a finite temperature.  Repeating this consideration for every channel,
we construct the phase diagram of the system in the ladder
approximation as a list of necessary conditions for the corresponding
instabilities:
\begin{mathletters}%
\label{LadderPhaseDiagram}
\begin{eqnarray}
   {\rm SSC:} & \quad & g_1+g_2<0, \\
   {\rm TSC:} & \quad & -g_1+g_2<0, \\
   {\rm CDW+:} & \quad & -2g_1+g_2-g_3>0, \\
   {\rm CDW-:} & \quad & -2g_1+g_2+g_3>0, \\
   {\rm SDW+:} & \quad & g_2+g_3>0, \\
   {\rm SDW-:} & \quad & g_2-g_3>0.
\end{eqnarray}
\end{mathletters}%

   The difference between the ladder and the parquet approximations
shows up when there are more than one appreciable bricks in the
problem.  Then, the vertex $f_{\rm SSC}$ contains not only the brick
$C_{\rm SSC}$, but other bricks as well, so Eqs.\ (\ref{NEWRG}) get
coupled.  This is the case, for example, for the 1D spinless
electrons, where the bricks $C$ and $Z$ are equally big, so they
cancel each other in $\gamma$ (see Sec.\ \ref{sec:spinless}).

\section{Results of Numerical Calculations}
\label{sec:numerical}

  The numerical procedure consists of three consecutive steps; each of
them involves solving differential equations by the fourth-order
Runge--Kutta method.  First, we solve parquet equations
(\ref{2Dgammas}) and (\ref{2Dbricks}) for the interaction vertices,
which are closed equations.  Then, we find the triangular vertices
${\cal T}_i$, whose equations (\ref{2DSCOP}) and (\ref{CSDWOPA})
involve the interaction vertices $\gamma_i$ through Eqs.\ (\ref{fSC}),
(\ref{fCDW}), and (\ref{fSDW}).  Finally, we calculate the
susceptibilities $\chi_i$ from Eqs.\ (\ref{2Dchii}), which depend on
the triangular vertices ${\cal T}_i$.

\begin{figure}
\centerline{\psfig{file=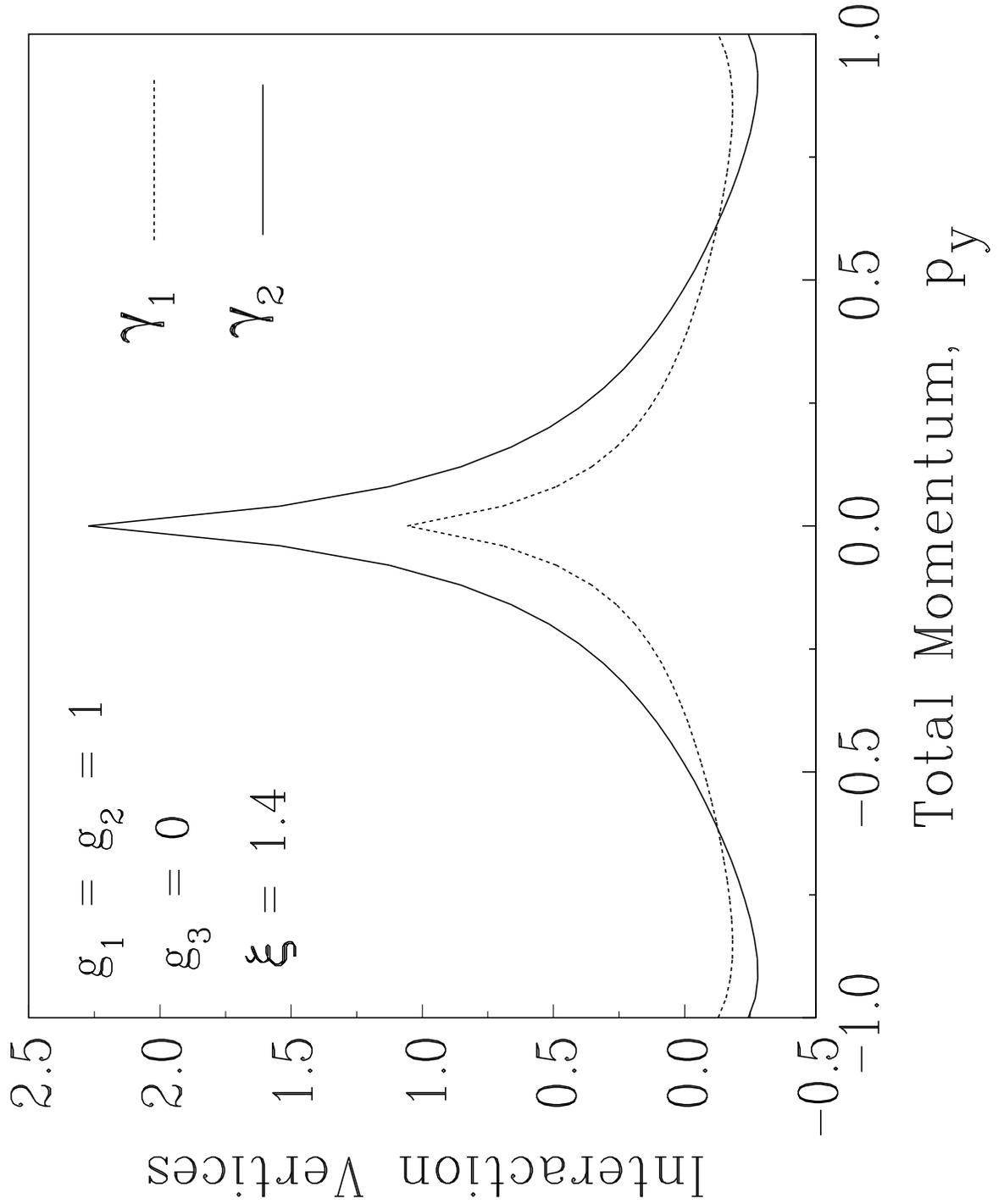,height=0.9\textheight}}
\caption{Interaction vertices
$\gamma_1(k_y^{(1)},k_y^{(2)};\:k_y^{(3)},k_y^{(4)};\:\xi)$ and
$\gamma_2(k_y^{(1)},k_y^{(2)};\:k_y^{(3)},k_y^{(4)};\:\xi )$ as
functions of the average momentum $p_y = (k_y^{(1)} + k_y^{(2)})/2$ of
the incoming electrons at $k_y^{(1)} = k_y^{(3)}$, $k_y^{(2)} =
k_y^{(4)}$, and $\xi = 1.4$.}
\label{fig:GammaData}
\end{figure}

\begin{figure}
\centerline{\psfig{file=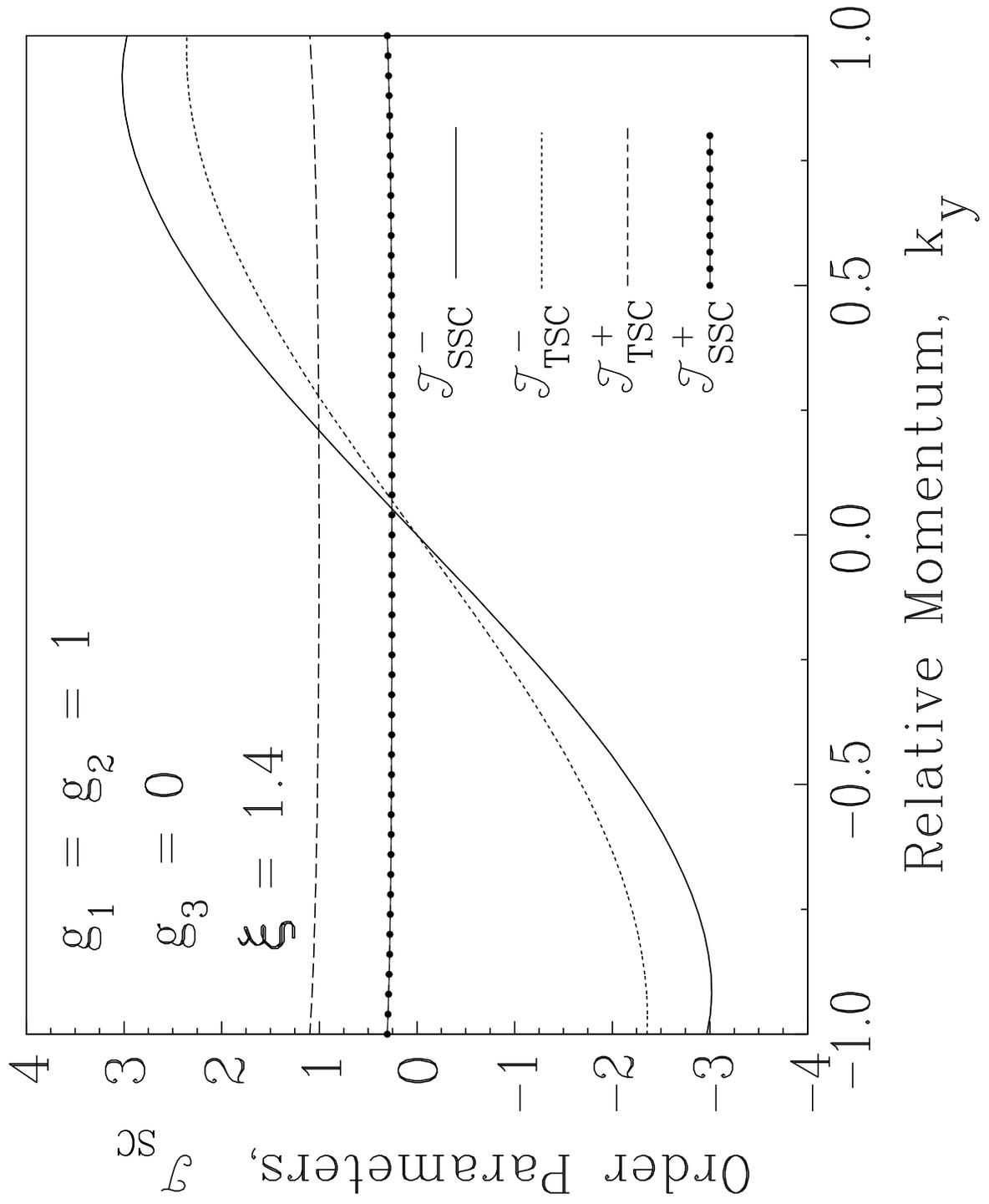,height=0.9\textheight}}
\caption{Superconducting order parameters ${\cal T}_{\rm SC}(k_y,q_y,
\xi)$ as functions of relative momentum $k_y$ at $q_y=0$ and
$\xi=1.4$.}
\label{fig:TSCData}
\end{figure}

\begin{figure}
\centerline{\psfig{file=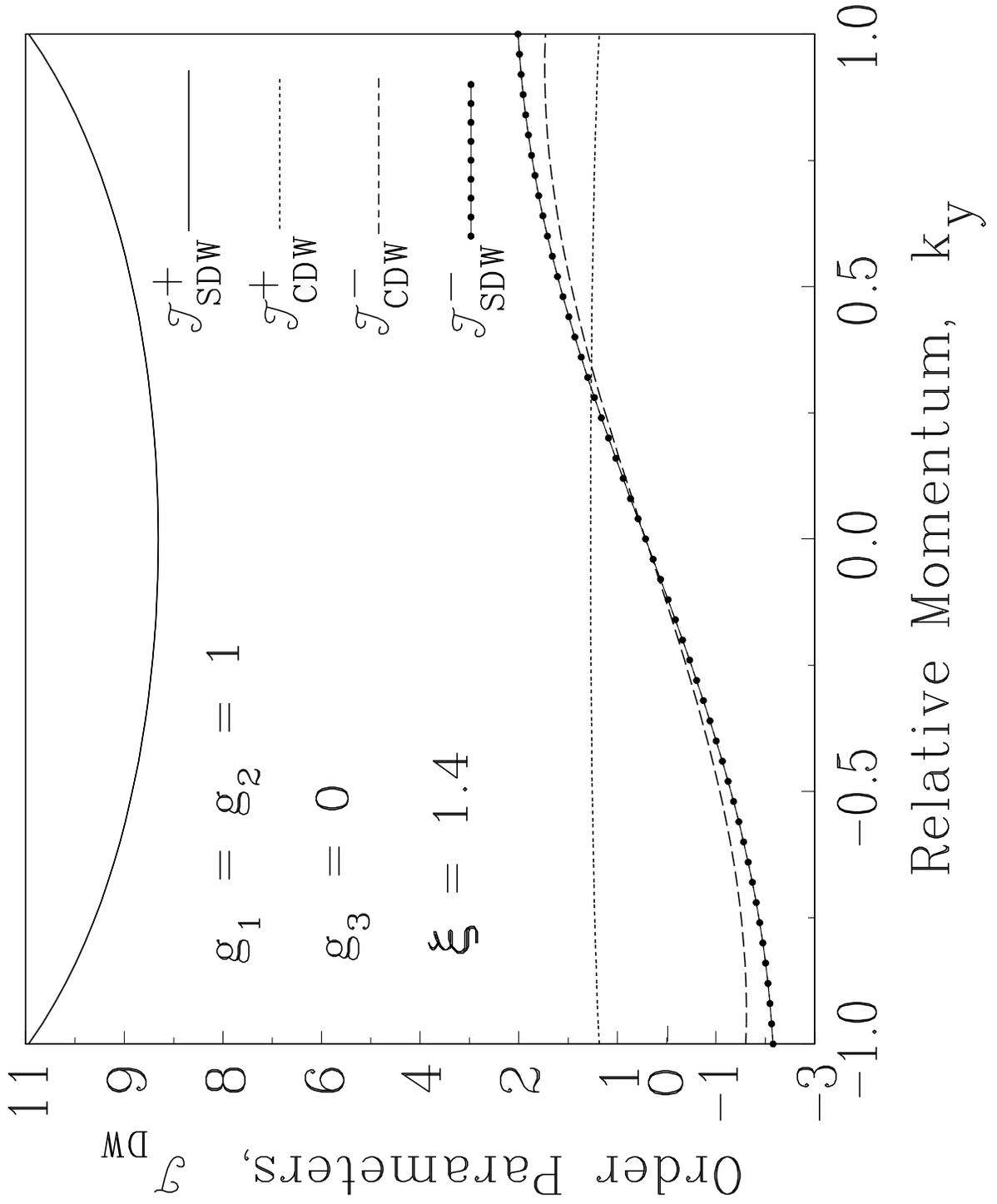,height=0.9\textheight}}
\caption{Density-wave order parameters ${\cal T}_{\rm DW}(k_y,q_y,
\xi)$ as the functions of relative momentum $k_y$ at $q_y=0$ and
$\xi=1.4$.}
\label{fig:TDWData}
\end{figure}

\begin{figure}
\centerline{\psfig{file=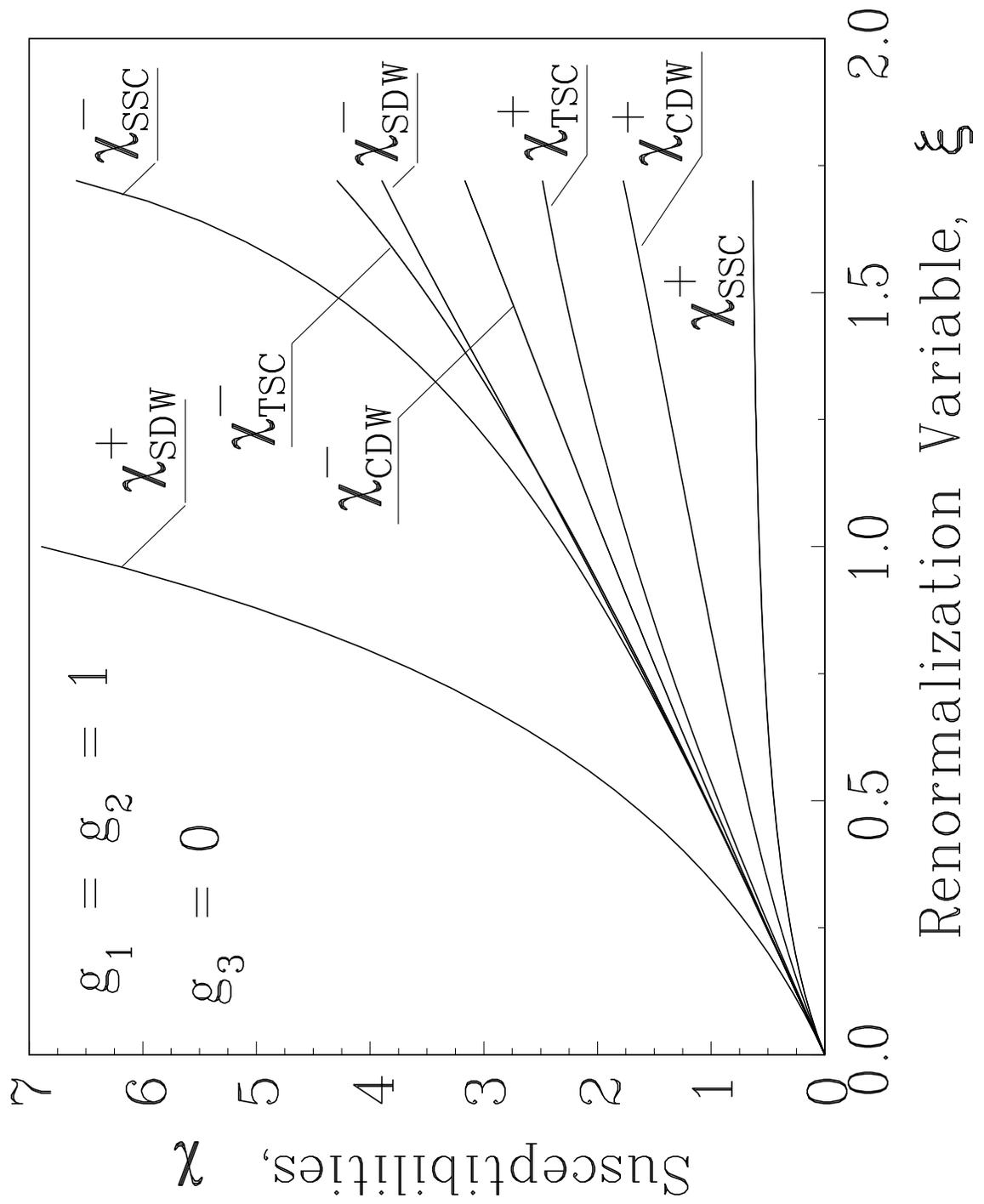,height=0.9\textheight}}
\caption{Evolution of susceptibilities $\chi_i(\xi)$ in the repulsive
Hubbard model without umklapp. $\chi_{\rm SDW}^+(\xi)$ diverges at
$\xi = \xi_{\rm SDW} = 1.76$.}
\label{fig:chi110}
\end{figure}

\begin{figure}
\centerline{\psfig{file=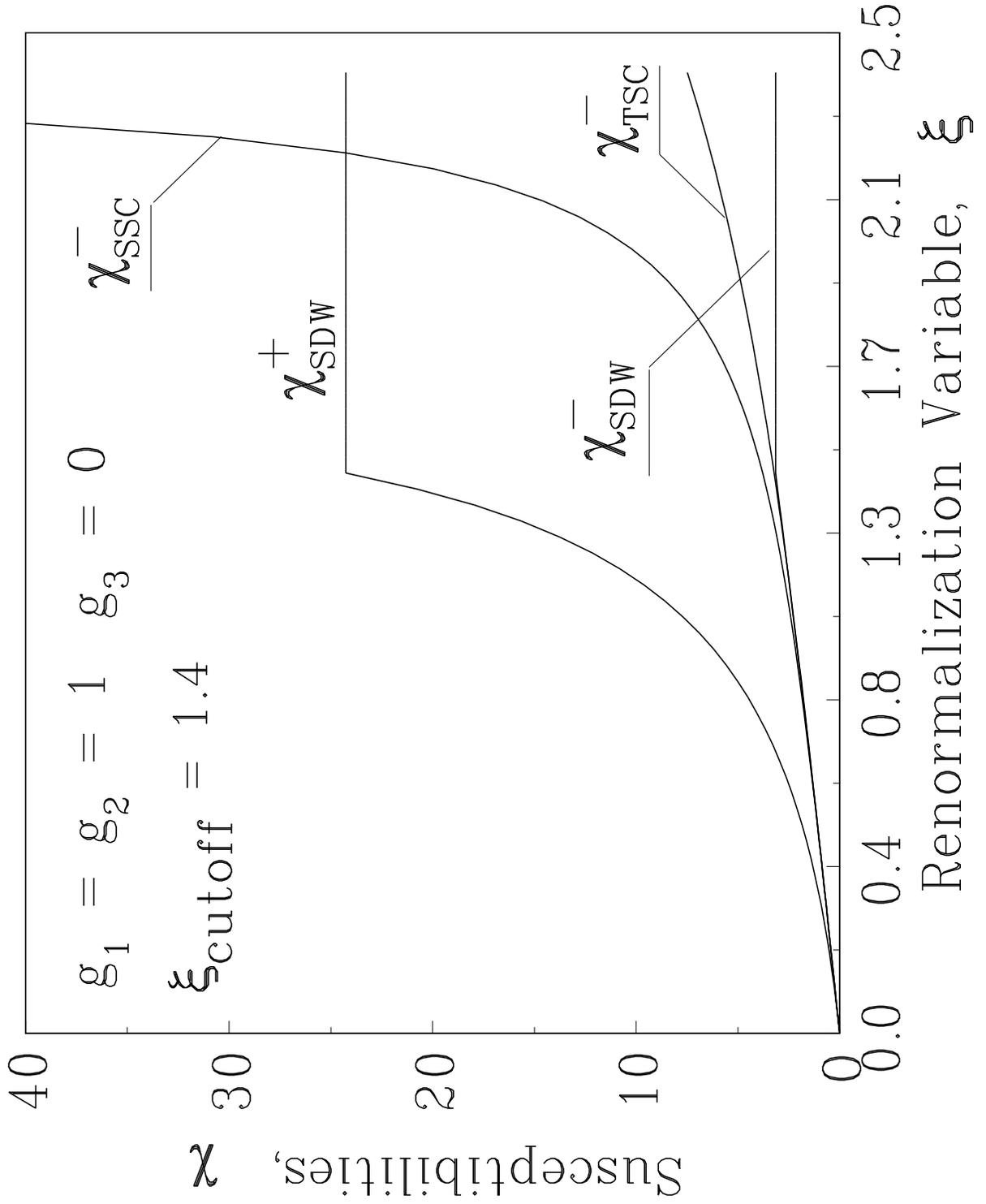,height=0.9\textheight}}
\caption{Evolution of the four leading susceptibilities in the Hubbard
model without umklapp, with the cutoff of all density-wave channels at
$\xi > \xi_{\rm cutoff} = 1.4$.  The susceptibility of antisymmetric
singlet superconductivity, $\chi_{\rm SSC}^-$, diverges at $\xi =
\xi_{\rm SSC}^- = 2.44$.}
\label{fig:chiCutoff}
\end{figure}

\begin{figure}
\centerline{\psfig{file=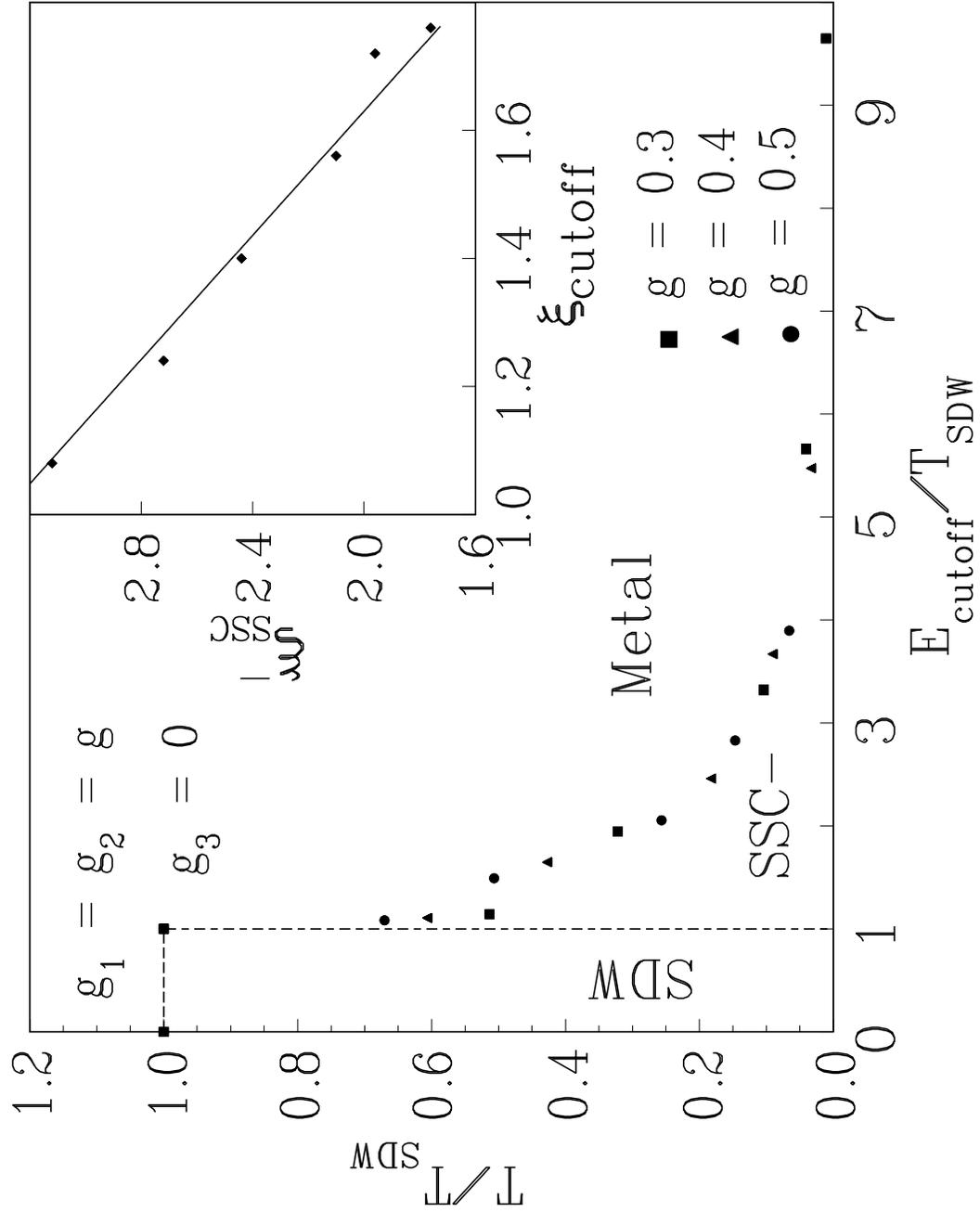,height=0.8\textheight}}
\caption{Phase diagram of the Hubbard model without umklapp
illustrates the dependence of the $d$-wave superconducting transition
temperature $T_{\rm SSC}^-$ on the cutoff energy $E_{\rm cutoff}$ for
different $g$.  The inset shows the same dependence in the logarithmic
variables $\xi_{\rm SSC}^-$ and $\xi_{\rm cutoff}$.  The solid line in
the inset is a fit: $\xi_{SSC}^- = a - b\; \xi_{\rm cutoff},\; a =
5.33,\; b = 2.06$.}
\label{fig:PhaseDiagram110}
\end{figure}

\begin{figure}
\centerline{\psfig{file=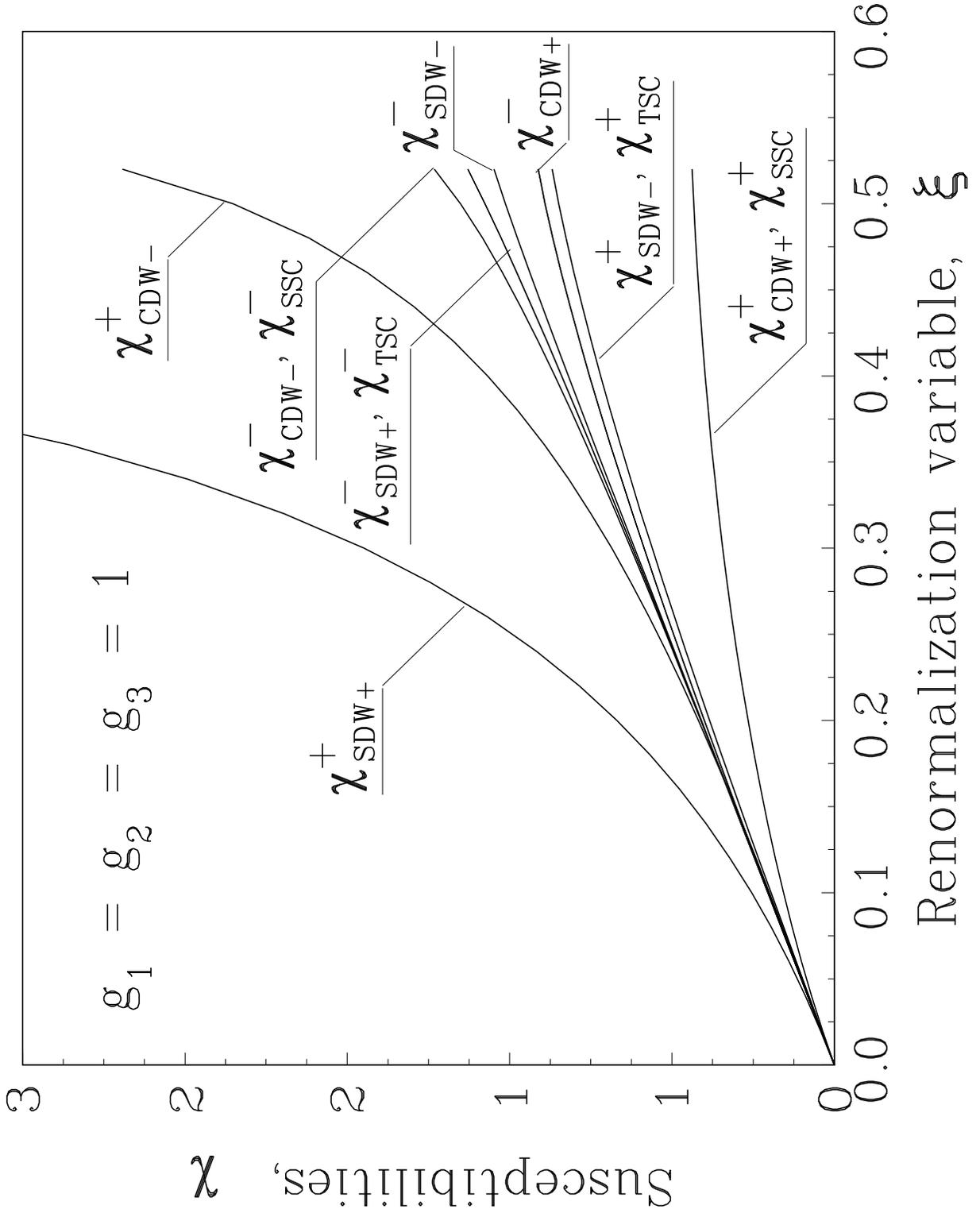,height=0.9\textheight}}
\caption{Evolution of generalized susceptibilities $\chi_i(\xi)$
in the Hubbard model with umklapp scattering. $\chi_{{\rm
SDW}+}^+(\xi)$ diverges at $\xi = \xi_{{\rm SDW}+}^+ = 0.54$.}
\label{fig:chi111}
\end{figure}

\begin{figure}
\centerline{\psfig{file=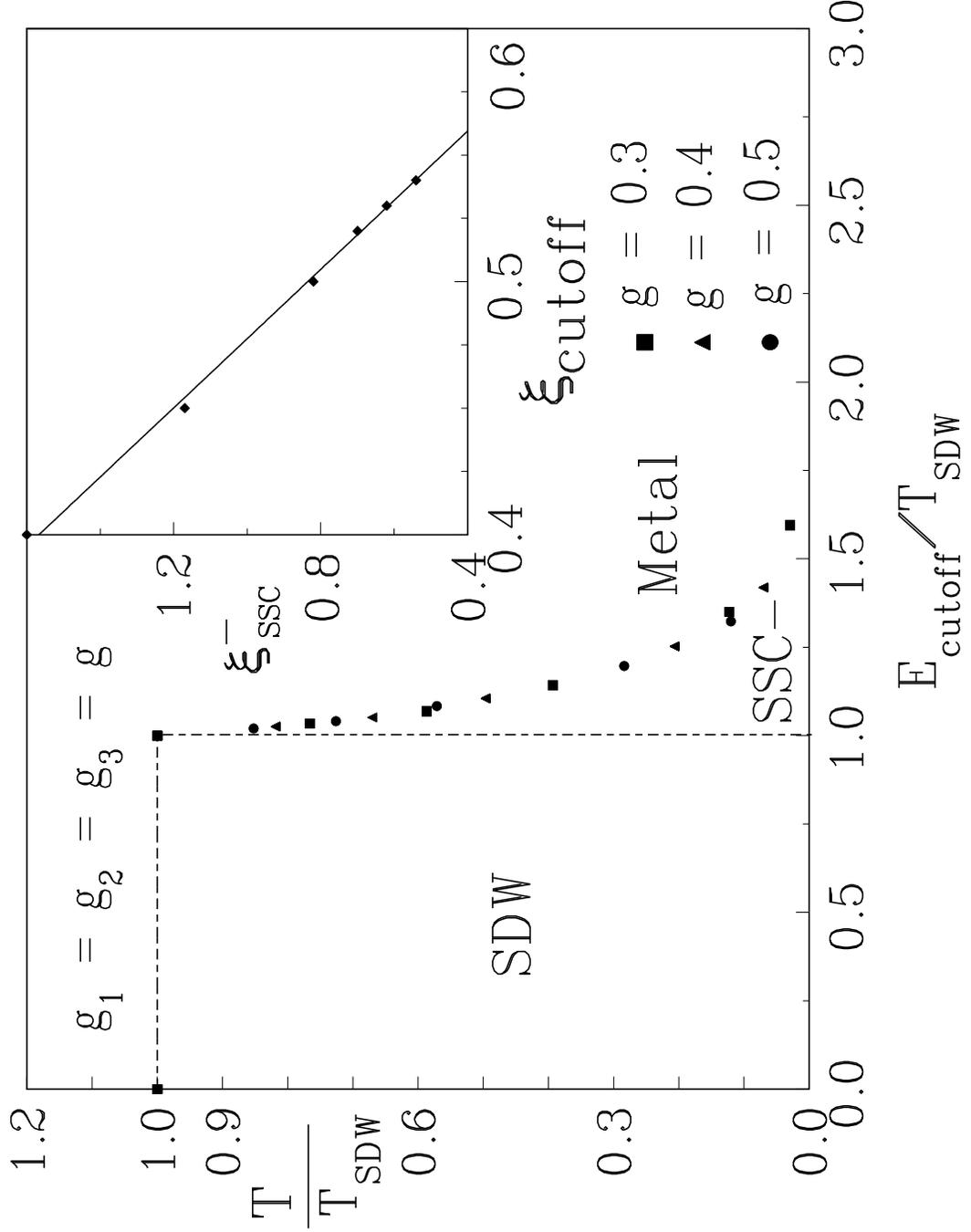,height=0.8\textheight}}
\caption{Phase diagram of the Hubbard model with umklapp scattering
illustrates the dependence of the critical temperature $T_{\rm SSC}^-$
on the cutoff energy $E_{\rm cutoff}$ for different $g$.  The inset
shows the dependence of $\xi_{\rm SSC}^-$ on $\xi_{\rm cutoff}$.  The
solid line in the inset is a fit: $\xi_{\rm SSC}^- = a - b\; \xi_{\rm
cutoff},\; a = 4.5,\; b = 7.32$.}
\label{fig:PhaseDiagram111}
\end{figure}

   We select the initial conditions for the interaction vertices to be
independent of the transverse momenta ${\cal K}$: $\gamma_i({\cal
K},\,\xi\!\!=\!\!0) = g_i$.  The momentum-independent interaction
naturally appears in the Hubbard model, where the interaction is local
in real space.  In this Chapter, the results are shown mostly for the
repulsive Hubbard model without umklapp: $g_1=g_2=g,\;g_3=0$ (Figs.\
\ref{fig:GammaData}--\ref{fig:PhaseDiagram110}), or with umklapp:
$g_1=g_2=g_3=g$ (Figs.\ \ref{fig:chi111}--\ref{fig:PhaseDiagram111}),
where $g$ is proportional the Hubbard interaction constant $U$.  The
absolute value of $g$ (but not the sign of $g$) is not essential in
our calculations, because it can be removed from the equations by
redefining $\xi$ to $\xi'=|g|\xi$.  After the redefinition, we
effectively have $|g|=1$ in the initial conditions.  The actual value
of $|g|$ matters only when the logarithmic variable $\xi'$ is
converted into the temperature according to the formula
$T=\mu\exp(-2\pi v_F\xi'/|g|)$.

   The initial independence of $\gamma_i({\cal K},\,\xi\!\!=\!\!0)$ on
${\cal K}$ does not imply that this property is preserved upon
renormalization.  On the contrary, during renormalization,
$\gamma_i({\cal K},\xi)$ develops a very strong dependence on ${\cal
K}$ and may even change sign in certain regions of the ${\cal
K}$-space.  We illustrate this statement in Fig.\ \ref{fig:GammaData}
by showing typical dependences of $\gamma_1({\cal K},\xi)$ and
$\gamma_2({\cal K},\xi)$ on the average momentum
$p_y=(k_y^{(1)}+k_y^{(2)})/2$ of the incoming electrons at $k_1=k_3$
and $k_2=k_4$ after some renormalization ($\xi = 1.4$).  In Figs.\
\ref{fig:GammaData}--\ref{fig:TDWData}, the upper and lower limits on
the horizontal axes are the boundaries $\pm k_y^{(0)}$ of the flat
region on the Fermi surface, which are set to $\pm1$ without loss of
generality.  One can observe in Fig.\ \ref{fig:GammaData} that the
electron-electron interaction becomes negative (attractive) at large
$p_y$, even though initially it was repulsive everywhere.

   Mathematically, the dependence of $\gamma_i({\cal K},\xi)$ on
${\cal K}$ arises because of the finite limits of integration,
$[-k_y^{(0)},k_y^{(0)}]$, imposed on the variables $k_y^{(A)}$ and
$k_y^{(B)}$ in Eqs.\ (\ref{2Dbricks}).  For example, in Eq.\
(\ref{C1}), when $p_y=(k_y^{(1)}+k_y^{(2)})/2$ equals zero,
$k_y^{(A)}$ may change from $-k_y^{(0)}$ to $k_y^{(0)}$ while
$k_y^{(B)}$ stays in the same interval.  However, when $p_y>0$,
$k_y^{(A)}$ has to be confined to a narrower interval
$[-k_y^{(0)}+2p_y,k_y^{(0)}]$ to ensure that
$k_y^{(B)}=2p_y-k_y^{(A)}$ stays within $[-k_y^{(0)},k_y^{(0)}]$.
This difference in the integration range subsequently generates the
dependence of $\gamma_i({\cal K},\xi)$ on $p_y$ and, more generally,
on ${\cal K}$. Since many channels with different geometrical
restrictions contribute to $\partial\gamma_i({\cal
K},\xi)/\partial\xi$ in Eqs.\ (\ref{2Dbricks}), the resultant
dependence of $\gamma_i({\cal K},\xi)$ on the four-dimensional vector
${\cal K}$ is complicated and hard to visualize.  Because of the
strong dependence of $\gamma_i({\cal K},\xi)$ on ${\cal K}$, it is not
possible to describe the 2D system by only three renormalizing charges
$\gamma_1(\xi)$, $\gamma_2(\xi)$, and $\gamma_3(\xi)$, as in the 1D
case.  Instead, it is absolutely necessary to consider an infinite
number of the renormalizing charges $\gamma_i({\cal K},\xi)$ labeled
by the continuous variable ${\cal K}$.  This important difference was
neglected in Ref.\ \cite{Marston93}, where the continuous variable
${\cal K}$ was omitted.

   Having calculated $\gamma_i({\cal K},\xi)$, we solve Eqs.\
(\ref{2DSCOP}) and (\ref{CSDWOPA}) for the triangular vertices (the
order parameters) ${\cal T}(k_y,q_y,\xi)$, which depend on both the
relative ($k_y$) and the total ($q_y$) transverse momenta.  We find
numerically that the order parameters with $q_y=0$ diverge faster than
those with $q_y\neq0$.  This is a natural consequence of the
integration range restrictions discussed above.  For this reason, we
discuss below only the order parameters with zero total momentum
$q_y=0$. We select the initial conditions for the symmetric and
antisymmetric order parameters in the form:
\begin{equation}
{\cal T}_i^+(k_y,\,\xi\!\!=\!\!0)=1,\quad
{\cal T}_i^-(k_y,\,\xi\!\!=\!\!0)=k_y.
\label{SA}
\end{equation}
In Figs.\ \ref{fig:TSCData} and \ref{fig:TDWData}, we present typical
dependences of the superconducting and density-wave order parameters
on the relative momentum $k_y$ at the same renormalization ``time''
$\xi = 1.4$ as in Fig.\ \ref{fig:GammaData}.  The singlet
antisymmetric component (${\cal T}_{\rm SSC}^{-}$) dominates among the
superconducting order parameters (Fig.\ \ref{fig:TSCData}), whereas
the symmetric SDW order parameter (${\cal T}_{SDW}^+$) is the highest
in the density-wave channel (Fig.\ \ref{fig:TDWData}).

   Having calculated the triangular vertices ${\cal T}$, we find the
susceptibilities from Eq.\ (\ref{2Dchii}).  The results are shown in
Fig.\ \ref{fig:chi110}.  The symmetric SDW has the fastest growing
susceptibility $\chi^+_{\rm SDW}$, which diverges at $\xi_{\rm
SDW}=1.76$.  This divergence indicates that a phase transition from
the metallic to the antiferromagnetic state takes place at the
transition temperature $T_{\rm SDW}=\mu\exp(-2\pi v_F\xi_{\rm
SDW}/g)$.  A similar result was obtained in Ref.\
\cite{Dzyaloshinskii72b} by analyzing the convergence radius of the
parquet series in powers of $g\xi$.  In the ladder approximation, the
SDW instability would take place at $\xi_{\rm SDW}^{\rm lad}=1/g_2=1$,
as follows from Eqs.\ (\ref{fSDW}) and (\ref{SW1}).  Since $\xi_{\rm
SDW}>\xi_{\rm SDW}^{\rm lad}$, the transition temperature $T_{\rm
SDW}$, calculated in the parquet approximation, is lower than the
temperature $T_{\rm SDW}^{\rm lad}$, calculated in the ladder
approximation: $T_{\rm SDW}<T_{\rm SDW}^{\rm lad}$.  The parquet
temperature is lower, because competing superconducting and
density-wave instabilities partially suppress each other.

   Thus far, we considered the model with ideally flat regions on the
Fermi surface.  Suppose now that these regions are only approximately
flat.  That is, they can be treated as being flat for the energies
higher than a certain value $E_{\rm cutoff}$, but a curvature or a
corrugation of the Fermi surface becomes appreciable at the smaller
energies $E<E_{\rm cutoff}$.  Because of the curvature, the Fermi
surface does not have nesting for $E<E_{\rm cutoff}$; thus, the
density-wave bricks in the parquet equations (\ref{2Dbricks}) stop to
renormalize.  Formally, this effect can be taken into account by
introducing a cutoff $\xi_{\rm cutoff}=(1/2\pi v_F)\ln(\mu/E_{\rm
cutoff})$, so that the r.h.s.\ of Eqs.\ (\ref{Z1})--(\ref{ZII}) for
the density-wave bricks are replaced by zeros at $\xi>\xi_{\rm
cutoff}$.  At the same time, Eqs.\ (\ref{C1}) and (\ref{C2}) for the
superconducting bricks remain unchanged, because the curvature of the
Fermi surface does not affect the superconducting instability with
$q_y=0$.  The change of the renormalization equations at $\xi_{\rm
cutoff}$ is not a completely rigorous way \cite{Luther88} to take into
account the Fermi surface curvature; however, this procedure permits
obtaining explicit results and has a certain qualitative appeal.  For
a more rigorous treatment of the corrugated Fermi surface problem see
Ref.\ \cite{Firsov}.

   In Fig.\ \ref{fig:chiCutoff}, we show the susceptibilities
calculated using the cutoff procedure with $\xi_{\rm cutoff}=1.4$.
The density-wave susceptibilities remain constant at $\xi>\xi_{\rm
cutoff}$.  At the same time, $\chi_{\rm SSC}^-(\xi)$ diverges at
$\xi_{\rm SSC}^-=2.44$ indicating a transition into the singlet
superconducting state of the $d$-wave type.  Thus, if the SDW
instability is suppressed, the system is unstable against formation of
the $d$-wave superconductivity. This result is in agreement with the
conclusions of Refs.\ \cite{Ruvalds95,Scalapino,Dzyaloshinskii87a}.

   From our numerical results, we deduce that the dependence of
$\xi_{\rm SSC}^-$ on $\xi_{\rm cutoff}$ is linear: $\xi_{\rm
SSC}^-=a-b\,\xi_{\rm cutoff}$ with $b=2.06$, as shown in the inset to
Fig.\ \ref{fig:PhaseDiagram110}.  Converting $\xi$ into energy in this
relation, we find a power law dependence:
\begin{equation}
    T_{\rm SSC}^- \propto \frac{1}{E_{\rm cutoff}^b}.
\label{TCR1}
\end{equation}
Eq.\ (\ref{TCR1}) demonstrates that increasing the cutoff energy
$E_{\rm cutoff}$ decreases the temperature of the superconducting
transition, $T_{\rm SSC}^-$.  Such a relation can be qualitatively
understood in the following way.  There is no bare interaction in the
superconducting $d$-wave channel in the Hubbard model, so the
transition is impossible in the ladder approximation.  The growth of
the superconducting $d$-wave correlations is induced by the growth of
the SDW correlations, because the two channels are coupled in the
parquet equations (\ref{NEWRG}).  If $E_{\rm cutoff}$ is high, the SDW
correlations do not have enough renormalization-group ``time'' $\xi$
to develop themselves because of the early cutoff of the density-wave
channels; thus, $T_{\rm SSC}^-$ is low.  Hence, decreasing $E_{\rm
cutoff}$ increases $T_{\rm SSC}^-$.  However, when $E_{\rm cutoff}$
becomes lower than $T_{\rm SDW}$, the SDW instability overtakes the
superconducting one.  Corresponding phase diagram is shown in Fig.\
\ref{fig:PhaseDiagram110}.  Generally speaking, the phase diagram
plotted in the energy variables, as opposed to the logarithmic
variables $\xi$, may depend on the absolute value of the bare
interaction constant $|g|$.  In Fig.\ \ref{fig:PhaseDiagram110}, we
placed the points for the several values of $g$ = 0.3, 0.4, and 0.5;
the phase boundary does not depend much on the choice of $g$.  The
phase diagram of Fig.\ \ref{fig:PhaseDiagram110} qualitatively
resembles the experimental one for the high-$T_c$ superconductors,
where transitions between the metallic, antiferromagnetic, and
superconducting states are observed.  The value of $E_{\rm cutoff}$
may be related to the doping level, which controls the shape of the
Fermi surface.  Taking into account the crudeness of our
approximations, detailed agreement with the experiment should not be
expected.

   We perform the same calculations also for the Hubbard model with
umklapp scattering ($g_1 = g_2 = g_3 =1$).  As one can see in Fig.\
\ref{fig:chi111}, where the susceptibilities are shown, the umklapp
process does not modify the qualitative picture. The leading
instability remains the SDW of the symmetric type, which is now also
symmetric with respect to the umklapp scattering, whereas the next
leading instability is the singlet $d$-wave superconductivity.  The
SDW has a phase transition at $\xi_{\rm SDW+}^+=0.54$, which is close
to the ladder result $\xi_{\rm SDW+}^{\rm lad}=1/(g_2+g_3)=0.5$.  Some
of the susceptibilities in Fig.\ \ref{fig:chi111} coincide exactly,
which is a consequence of a special SU(2)$\times$SU(2) symmetry of the
Hubbard model at the half filling \cite{SO(4)}.  The phase diagram
with the energy cutoff (Fig.\ \ref{fig:PhaseDiagram111}) is similar to
the one without umklapp (Fig.\ \ref{fig:PhaseDiagram110}), but the
presence of the umklapp scattering decreases the transition
temperature of the $d$-wave superconductivity.

\begin{figure}
\centerline{\psfig{file=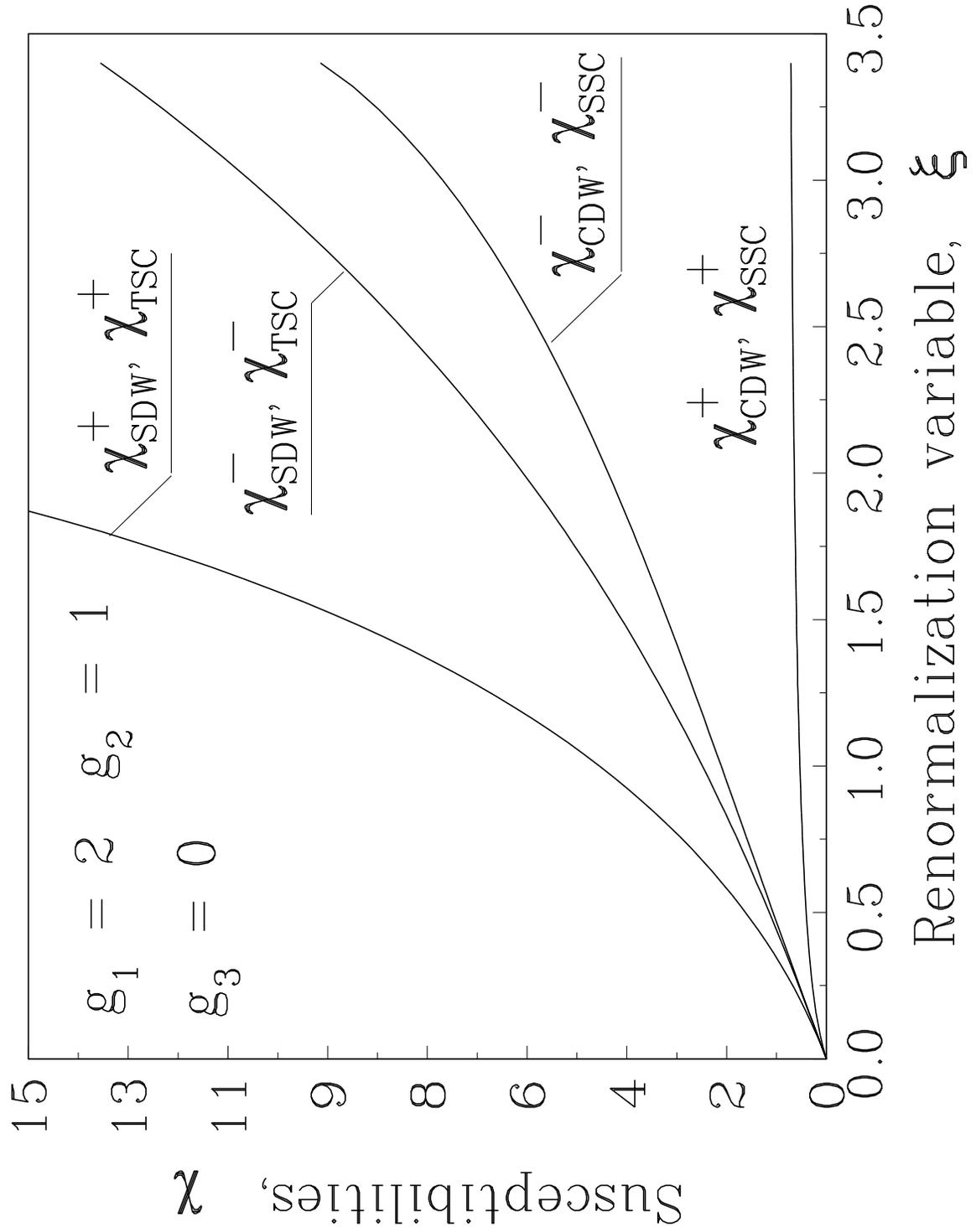,height=0.9\textheight}}
\caption{Evolution of susceptibilities in the case where $g_1 = 2$,
$g_2 = 1$, and $g_3 = 0$. $\chi_{\rm SDW}^+(\xi)$ and $\chi_{\rm
TSC}^-(\xi)$ diverge at $\xi = 3.48$.}
\label{fig:chi210}
\end{figure}

   An important issue in the study of the 1D electron gas is the
so-called $g$-ology phase diagram, which was constructed for the first
time by Dzyaloshinskii and Larkin \cite{Dzyaloshinskii72a}.  They
found that, in some regions of the $(g_1,g_2,g_3)$ space, the 1D
electron system develops a charge or spin gap, which is indicated by
divergence of $\gamma_i(\xi)$ with increasing $\xi$.  In the region
where none of the gaps develops, the Luttinger liquid exists.  It is
interesting whether such a region may exist in our 2D model.  To study
the phase diagram of the 2D system, we repeat the calculations,
systematically changing relative values of $g_1$, $g_2$, and $g_3$.
From the physical point of view, the relative difference of $g_1$,
$g_2$, and $g_3$ roughly mimics dependence of the interaction vertex
on the momentum transfer.  As an example, we show the susceptibilities
in the case where $g_1=2$, $g_2=1$, and $g_3=0$ in Fig.\
\ref{fig:chi210}.  In this case, the leading instabilities are
simultaneously the triplet superconductivity of the symmetric type
(TSC+) and the spin-density wave.

   For all studied sets of $g_i$, we find that the leading
instabilities calculated in the parquet and the ladder approximations
always coincide.  (We do not introduce the energy cutoff here.)  Thus,
the parquet effects do not modify the $g$-ology phase diagram of the
2D model derived in the ladder approximation, even though the
transition temperatures in the parquet approximation are always lower
than those obtained in the ladder approximation.  In that sense, the
parquet corrections are much less important in the 2D case than in the
1D case.  From the mathematical point of view, this happens because a
leading divergent brick develops a strong dependence on the transverse
momenta ${\cal K}$ and acquires the so-called mobile pole structure
\cite{Gorkov74,Brazovskii71,Dzyaloshinskii72b}:
\begin{equation}
   Z({\cal K},\xi)\propto\frac{1}{\xi_c({\cal K})-\xi}.
\label{MovingPole}
\end{equation}
The name ``mobile pole'' is given, because the position of the pole in
$\xi$ in Eq.\ (\ref{MovingPole}), $\xi_c({\cal K})$, strongly depends
on the momenta ${\cal K}$. It was shown in Refs.\
\cite{Brazovskii71,Gorkov74,Dzyaloshinskii72b} that, because of the
mobility of the pole, the leading channel decouples from the other
channels, and the parquet description effectively reduces to the
ladder one, as described at the end of Sec.\ \ref{sec:2D}.  The phase
diagram of the 2D system in the ladder approximation is given by Eqs.\
(\ref{LadderPhaseDiagram}).  It follows from Eqs.\
(\ref{LadderPhaseDiagram}) that every point in the $(g_1,g_2,g_3)$
space has some sort of instability.  Thus, the Luttinger liquid,
defined as a nontrivial metallic ground state where different
instabilities mutually cancel each other, does not exist in the 2D
model.

   Generally speaking, other models may have different types of
solutions of the fast parquet equations, such as immobile poles
\cite{Gorkov74} or a self-similar solution \cite{Yakovenko93a}, the
latter indeed describing some sort of a Luttinger liquid.  In our
study of a 2D model with the van Hove singularities
\cite{Dzyaloshinskii87a}, we found a region in the $g$-space without
instabilities, where the Luttinger liquid may exist
\cite{Dzyaloshinskii}.  However, we find only the mobile-pole
solutions in the present 2D model.

\section{Conclusions}
\label{sec:conclusion}

   In this paper we derive and numerically solve the parquet equations
for the 2D electron gas whose Fermi surface contains flat regions.
The model is a natural generalization of the 1D electron gas model,
where the Luttinger liquid is known to exist.  We find that, because
of the finite size of the flat regions, the 2D parquet equations
always develop the mobile pole solutions, where the leading
instability effectively decouples from the other channels.  Thus, a
ladder approximation is qualitatively (but not necessarily
quantitatively) correct for the 2D model, in contrast to the 1D case.
Whatever the values of the bare interaction constants are, the 2D
system always develops some sort of instability.  Thus, the Luttinger
liquid, defined as a nontrivial metallic ground state where different
instabilities mutually cancel each other, does not exist in the 2D
model, contrary to the conclusions of Refs.\
\cite{Mattis87,Hlubina94}.

   In the case of the repulsive Hubbard model, the leading instability
is the SDW, i.e., antiferromagnetism \cite{Dzyaloshinskii72b}.  If the
nesting of the Fermi surface is not perfect, the SDW correlations do
not develop into a phase transition, and the singlet superconductivity
of the $d$-wave type appears in the system instead.  These results may
be relevant for the high-$T_c$ superconductors and are in qualitative
agreement with the findings of Refs.\
\cite{Ruvalds95,Scalapino,Dzyaloshinskii87a}.

   In the bosonization procedure
\cite{Haldane92,Khveshchenko93a,Khveshchenko94b,Marston93,Marston,Fradkin,LiYM95,Kopietz95},
a higher-dimensional Fermi surface is treated as a collection of flat
patches. Since the results of our paper do not depend qualitatively on
the size of the flat regions on the Fermi surface, the results may be
applicable, to some extent, to the patches as well. Precise relation
is hard to establish because of the infinitesimal size of the patches,
their different orientations, and uncertainties of connections between
them. On the other hand, the bosonization procedure seems to be even
better applicable to a flat Fermi surface, which consists of a few big
patches. Mattis \cite{Mattis87} and Hlubina \cite{Hlubina94} followed
that logic and claimed that the flat Fermi surface model is exactly
solvable by the bosonization and represents a Luttinger liquid. The
discrepancy between this claim and the results our paper indicates
that some conditions must restrict the validity of the bosonization
approximations. Luther gave a more sophisticated treatment to the flat
Fermi surface problem by mapping it onto multiple quantum chains
\cite{Luther94}. He found that the bosonization converts the
interaction between electrons into the two types of terms, roughly
corresponding to the two terms of the sine-Gordon model: the
``harmonic'' terms $(\partial \varphi/\partial x)^2$ and the
``exponential'' terms $\exp(i\varphi)$, where $\varphi$ is a
bosonization phase. The harmonic terms can be readily diagonalized,
but the exponential terms require a consistent renormalization-group
treatment. If the renormalization-group equations were derived in the
bosonization scheme of \cite{Luther94}, they would be the same as the
parquet equations written in our paper, because the
renormalization-group equations do not depend on whether the boson or
fermion representation is used in their derivation \cite{Wiegmann78}.

   Long time ago, Luther bosonized noninteracting electrons on a
curved Fermi surface \cite{Luther79}; however, the interaction between
the electrons remained intractable because of the exponential
terms. The recent bosonization in higher dimensions
\cite{Haldane92,Khveshchenko93a,Khveshchenko94b,Marston93,Marston,Fradkin,LiYM95,Kopietz95}
managed to reformulate the problem in the harmonic terms only. This is
certainly sufficient to reproduces the Landau description of sound
excitations in a Fermi liquid \cite{Landau-IX}; however, it may be not
sufficient to derive the electron correlation functions. The validity
of the harmonic approximation is hard to trace for a curved Fermi
surface, but considerable experience has been accumulated for the flat
Fermi surface models.

   In the model of multiple 1D chains without single-electron
tunneling between the chains and with forward scattering between
different chains, the bosonization produces the harmonic terms only,
thus the model can be solved exactly
\cite{Larkin73,Gutfreund76b}. However, a slight modification of the
model by introducing backward scattering between different chains
\cite{Gorkov74,PALee77} or interaction between four different chains
\cite{Yakovenko87} adds the exponential terms, which destroy the exact
solvability and typically lead to a CDW or SDW instability. Even if no
instability occurs, as in the model of electrons in a high magnetic
field \cite{Yakovenko93a}, the fast parquet method shows that the
electron correlation functions have a complicated, nonpower structure,
which is impossible to obtain within the harmonic bosonization.
Further comparison of the fast parquet method and the bosonization in
higher dimensions might help to establish the conditions of
applicability of the two complementary methods.

   The work at Maryland was partially supported by the NSF under Grant
DMR--9417451, by the Alfred P.~Sloan Foundation, and by the David and
Lucile Packard Foundation.

\newpage

\newpage

\newpage

\newpage

\newpage

\newpage

\newpage

\newpage

\newpage

\newpage

\newpage

\newpage

\newpage

\newpage

\newpage

\newpage

\newpage

\newpage

\newpage

\end{document}